\documentclass[aps,preprint,nofootinbib,preprintnumbers,eqsecnum,superscriptaddress]{revtex4-1}

\usepackage{color}

\newcommand{\nb}[1]{\color{blue}}

\newcommand{\HL}[1]{{\bf \textcolor{magenta}{#1}}}
\newcommand{\hl}[1]{\color{magenta}}

\usepackage[
	  pagebackref=false,
	  colorlinks=true,
      linkcolor=blue,
      urlcolor=blue,
      filecolor=black,
      citecolor=red,
      pdfstartview=FitV,
      pdftitle={},
        pdfauthor={Paolo Glorioso, Michael Crossley, Hong Liu},
        pdfsubject={},
        pdfkeywords={},
        pdfpagemode=None,
        bookmarksopen=true
      ]{hyperref}

\usepackage[normalem]{ulem}
\usepackage{amsmath}
\usepackage{enumerate}
\usepackage{amsfonts}
\usepackage{epsfig}
\usepackage{mathbbol}

\def\Tr{\mathop{\rm Tr}}
\def\tr{\mathop{\rm tr}}

\newcommand\half{{\ensuremath{\frac{1}{2}}}}
\newcommand\p{\ensuremath{\partial}}

\newcommand{\be}{\begin{equation}}
\newcommand{\ee}{\end{equation}}
\newcommand{\bea}{\begin{eqnarray}}
\newcommand{\eea}{\end{eqnarray}}
\newcommand{\bega}{\begin{gather}}
\newcommand{\eega}{\end{gather}}

\newcommand{\bi}{\begin{itemize}}
\newcommand{\ei}{\end{itemize}}
\newcommand{\ben}{\begin{enumerate}}
\newcommand{\een}{\end{enumerate}}
\newcommand{\bca}{\begin{cases}}
\newcommand{\eca}{\end{cases}}
\newcommand{\bln}{\begin{align}}
\newcommand{\eln}{\end{align}}
\newcommand{\bst}{\begin{split}}
\newcommand{\est}{\end{split}}
\def\ie{\begin{equation}\begin{aligned}}
\def\fe{\end{aligned}\end{equation}}
\newcommand{\bma}{\le(\begin{matrix}}
\newcommand{\ema}{\end{matrix}\ri)}

\newcommand\al{{\alpha}}
\newcommand\ep{\epsilon}
\newcommand\sig{\sigma}
\newcommand\Sig{\Sigma}
\newcommand\lam{\lambda}
\newcommand\Lam{\Lambda}

\newcommand\de{{\ensuremath{{\delta}}}}
\newcommand\De{{\ensuremath{{\Delta}}}}
\newcommand\vp{\varphi}

\newcommand\vep{\varepsilon}

\newcommand\da{{\dagger}}
\newcommand\nab{{\nabla}}
\newcommand\Th{{\Theta}}
\def\th{{\theta}}

\newcommand\ov{\over}
\newcommand\ha{{\half}}

\def\le{\left}
\def\ri{\right}

\newcommand\sC{{\ensuremath{{\mathcal C}}}}

\newcommand\sE{{\ensuremath{{\mathcal E}}}}

\newcommand\sI{{\ensuremath{{\mathcal I}}}}
\newcommand\sG{{\ensuremath{{\mathcal G}}}}

\newcommand\sK{{\ensuremath{{\mathcal K}}}}
\newcommand\sL{{\ensuremath{{\mathcal L}}}}

\newcommand\sN{{\ensuremath{{\mathcal N}}}}

\newcommand\sP{{\ensuremath{{\mathcal P}}}}
\newcommand\sQ{{\mathcal Q}}

\newcommand\sJ{{\mathcal J}}

\newcommand\sS{{\mathcal S}}
\newcommand\sT{{\mathcal T}}

\newcommand\vx{{\vec x}}

\newcommand{\hmu}{{\hat \mu}}

\newcommand{\fb}{{\mathfrak{b}}}
\newcommand{\fc}{{\mathfrak{c}}}

\newcommand{\ft}{{\mathfrak t}}

\begin{document}

%\title{A covariant formulation of Local KMS symmetry for fluctuating hydrodynamics}
\title{Effective field theory for dissipative fluids (II): \\
classical limit, dynamical KMS symmetry and entropy current}

\preprint{MIT-CTP/4860, EFI/17-2}

\author{Paolo Glorioso}
\affiliation{Kadanoff Center for Theoretical Physics and Enrico Fermi Institute\\
University of Chicago, Chicago, IL 60637, USA}

\author{Michael Crossley}
\affiliation{Center for Theoretical Physics, \\
Massachusetts
Institute of Technology,
Cambridge, MA 02139 }

\author{Hong Liu}
\affiliation{Center for Theoretical Physics, \\
Massachusetts
Institute of Technology,
Cambridge, MA 02139 }

\begin{abstract}

\noindent  In this paper we further develop the fluctuating hydrodynamics proposed  in~\cite{CGL} in a number of ways.
We first work out in detail the classical limit of the hydrodynamical action, which exhibits many simplifications.
 In particular, this enables a transparent formulation of the action in physical spacetime in the presence of
arbitrary external fields. It also helps to clarify issues related to field redefinitions and frame choices.
 We then propose that the action is invariant under a $Z_2$ symmetry to which we refer as the dynamical KMS symmetry.
The dynamical KMS symmetry is physically equivalent  to the previously proposed local KMS condition in the classical limit,  but is more convenient to implement and more general. It is applicable to any states  in local equilibrium rather than just thermal density matrix perturbed by external background fields.
 Finally we elaborate the formulation for a conformal fluid, which contains some new features, and work out the explicit form
 of the entropy current to second order in derivatives for a neutral conformal fluid.

\end{abstract}

\today

\maketitle

\tableofcontents

\section{Introduction}

For a quantum many-body system in local thermal equilibrium, in generic situations, the only long-lived modes are those associated with conserved quantities such as energy-momentum tensor and conserved currents for some global symmetries.   Recently, using this as a starting point we proposed a new formulation of
fluctuating hydrodynamics as a universal low energy effective theory of a quantum many-body system at a finite temperature~\cite{CGL}.\footnote{For other recent discussions of fluctuating hydrodynamics, see~\cite{Dubovsky:2005xd,Dubovsky:2011sj,Endlich:2012vt,Dubovsky:2011sk,Endlich:2010hf,
Nicolis:2011ey,Nicolis:2011cs,Delacretaz:2014jka,Geracie:2014iva,Grozdanov:2013dba,
Kovtun:2014hpa,Harder:2015nxa,
Haehl:2015foa,Haehl:2015uoc,Haehl:2015pja,Haehl:2015foa,Haehl:2015uoc,Haehl:2016pec,
Andersson:2013jga,Floerchinger:2016gtl} and in holographic
context~\cite{Nickel:2010pr,deBoer:2015ija,Crossley:2015tka}.} The theory gives a path integral formulation of hydrodynamics  which systematically incorporates effects of fluctuations, including nonlinear interactions involving noises as well as non-equilibrium fluctuation-dissipation relations.
The conventional hydrodynamical equations of motion are recovered as saddle point equations, and the stochastic hydrodynamics is recovered by truncating the noise part of the action to quadratic level.

We now summarize the salient aspects of the theory of~\cite{CGL}.
 For definiteness we will consider a system with a $U(1)$ global symmetry.
Consider the closed time path (CTP) generating functional
for the stress tensor and  $U(1)$ current
in a thermal density matrix $\rho_0$
\bea \label{pager1}
e^{W  [ g_{1\mu \nu} , A_{1\mu}; g_{2 \mu \nu}, A_{2 \mu}] }
&\equiv&  \Tr \le[U (+\infty, -\infty; g_{1\mu \nu}, A_{1\mu}) \rho_0 U^\da (+\infty, -\infty; g_{2\mu \nu}, A_{2 \mu})\ri] \\
& = & \int D \chi_1 D \chi_2 \, e^{i I_{\rm hydro} [\chi_1, g_1, A_1 ; \chi_2,g_2,A_2]}
\label{path}
\eea
where $U (t_2, t_1; g_{\mu \nu}, A_{\mu})$ is the evolution operator of the system from $t_1$ to $t_2$
in the presence of  a spacetime metric $g_{\mu \nu}$ (sources for stress tensor) and an external vector field $A_\mu$ (sources for the $U(1)$ current). The sources are taken to be slowly varying functions and there are two copies of them, one for each leg of the CTP contour.
%The second line~\eqref{fpth} is the ``microscopic'' path integral description, with $\psi_{1,2}$  denoting dynamical variables for the two copies of spacetime of the CTP and $S_0 [\psi]$ the microscopic action.
The second line~\eqref{path} should be imagined as obtained by integrating out all the fast modes of the system
with $\chi_{1,2}$ denoting the remaining slow modes (hydrodynamical modes), and $I_{\rm hydro}$ is the effective action for them.
Again there are two copies of hydrodynamical modes. While in practice the integrations from~\eqref{pager1} to~\eqref{path} cannot be done, $I_{\rm hydro}$ can be obtained as the most general local action once we have identified  the dynamical variables $\chi_{1,2}$ and the symmetries $I_{\rm hydro}$ should obey.

 %(two copies of) hydrodynamic modes and Note that while in~\eqref{fpth} $\rho_0$ is encoded in the boundary conditions of the path integrals,  in~\eqref{path} $\rho_0$ is encoded in the coefficients of effective action $I_{\rm hydro}$.

% \begin{figure}[!h]
%\begin{center}
%$
%\begin{array}{cc}
%\includegraphics[scale=1]{fig2b.pdf} \quad
%\end{array}
%$
%\end{center}
%\caption{ Closed time path contour with initial density matrix $\rho_0$: The two lines represent two copies of spacetime.  Inserted operators should be path ordered as indicated by the arrows. \HL{Mentioning external fields.}}
% \label{fig:SK}
%\end{figure}

\begin{figure}[!h]
\begin{center}
\includegraphics[scale=0.8]{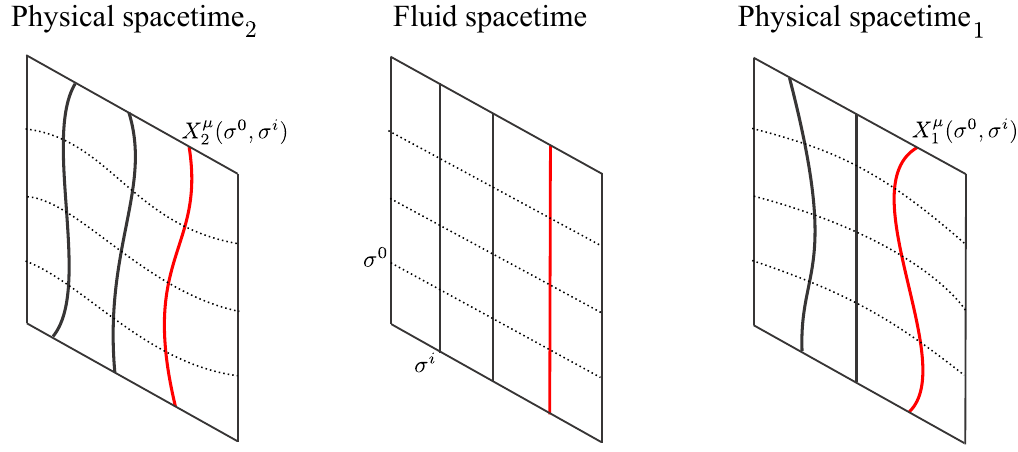}
\end{center}
\caption{Relations between the fluid spacetime and two copies of physical spacetimes.  The red straight
line in the fluid spacetime with constant $\sig^i$ is mapped by $X^\mu_{1,2} (\sig^0, \sig^i)$ to
physical spacetime trajectories (also in red) of the  corresponding fluid element.
 }
\label{fig:Lagrange}
\end{figure}

In~\cite{CGL} we developed an ``integrating-in'' procedure to identify the slow degrees of freedom associated with a
conserved quantity. For the stress tensor this leads to a doubled
version of the Lagrange description of fluid flows, with the corresponding $\chi_{1,2}$ given by
mappings $X^\mu_{1,2} (\sig^a)$ between a ``fluid spacetime," whose coordinates $\sig^a$ label fluid elements and their internal clocks, and the two copies of physical spacetimes with coordinates $X^\mu_{1,2}$ respectively. See Fig.~\ref{fig:Lagrange}.
The slow degrees of freedom for the $U(1)$ current are  $\vp_{1,2} (\sig^a)$  which can be interpreted as $U(1)$ phase rotations in two physical spacetimes associated for each fluid elements.\footnote{Other recent work which uses effective action from CTP to describe fluctuating hydrodynamics include~\cite{Grozdanov:2013dba,Haehl:2015pja,Haehl:2015foa,Haehl:2015uoc}. In particular similar variables were also used in~\cite{Grozdanov:2013dba,Kovtun:2014hpa,Harder:2015nxa,Haehl:2015foa,Haehl:2015pja,Haehl:2015uoc}.}
One also needs to introduce an additional scalar field $\tau (\sig^a)$ which gives the local proper temperature in fluid spacetime
\be \label{tay0}
T (\sig) = {1 \ov \beta(\sig)} =  T_0 e^{-\tau (\sig)} \ .
\ee
$T_0 = {1 \ov \beta_0}$ is the temperature at infinities where we take all external sources and dynamical fields to vanish. Note that there is only one $\tau$ field rather than two copies.
The standard variables such as local velocity and chemical potential
are built from symmetric combinations of $X^\mu_{1,2} , \vp_{1,2}$, while their antisymmetric combinations can be interpreted as
corresponding ``noises.''

It turns out these variables are not enough. In order to ensure the unitarity of~\eqref{pager1}
 one needs to introduce  anti-commuting partners (``ghosts'') for dynamical variables and require the full action to satisfy a BRST-type fermionic symmetry.\footnote{The need for anti-commuting degrees of freedom and BRST symmetry in path integral formulation of stochastic systems has been well known since 1970's. See e.g.~\cite{zinnjustin} for a review.
 Their presence for fluctuating hydrodynamics has also been discussed recently in~\cite{Haehl:2015foa,Haehl:2015uoc,Haehl:2016pec}.}
  %Interestingly, these fermionic degrees of freedom survive in the classical limit.  In other words, for a classical statistical system which can arise from a limit of a quantum system,  fermionic variables are needed to properly describe its thermal fluctuations.
  In this paper we will focus on the bosonic part and so will suppress all ghost variables below.
 The structure with ghost sector is elucidated in~\cite{ping,yarom}.

%Thus the full set of dynamical variables are $X_{1,2}^\mu (\sig^a), \vp_{1,2} (\sig^a), \tau(\sig^a)$
%all of which are scalar fields in the fluid spacetime.
%Interestingly, to ensure the unitarity condition~\eqref{top1} one also needs to introduce an anti-commuting partner for each bosonic degree of freedom and requires the action to be invariant under a BRST-type fermionic symmetry.

%For hydrodynamics we are interested in correlation functions of the stress tensor $T^{\mu \nu}$ and conserved current(s) $J^\mu$ for some global symmetry(ies) (if there is any). The corresponding sources are spacetime metric $g_{\mu \nu}$ and ``gauge field'' $A_\mu$. Thus $\phi$ in~\eqref{odd}--\eqref{1kms} should be understood as short-hand notation for the collection $\{g_{\mu \nu}, A_\mu \}$.
%At low energies and long distances (i.e. with background fields $g_{1} , A_1, g_{2}, A_2$ all slowly varying functions of spacetime),

In terms of  variables described above~\eqref{path} can be written more explicitly as
\be \label{qft0}
e^{W [ g_{1} , A_1; g_{2}, A_2]} = \int D X_1 D X_2  D\vp_1 D \vp_2 D \tau   \, e^{i  I_{\rm hydro} [h_{1},  B_1; h_2, B_{2}; \tau]},
\ee
where ($s=1,2$ and no summation over $s$)
\bega \label{hdef1}
h_{sab} (\sig)  =  {\p X^\mu_s \ov \p \sig^a} g_{s\mu \nu} (X_s (\sig)) {\p X_s^\nu \ov \p \sig^b} , \qquad
B_{sa} (\sig) = {\p X^\mu_s \ov \p \sig^a}  A_{s\mu} (X_s (\sig)) + \p_a \vp_s (\sig) \ .
\end{gather}
$h_{1,2}$ are  pull-backs of the spacetime metrics to the fluid spacetime and similarly the first terms in $B_{1,2}$ are  pull-backs of spacetime vector sources. Due to conservation of the stress tensor and $U(1)$ current, $W[ g_{1} , A_1; g_{2}, A_2]$ should be invariant under independent diffeomorphisms of the two copies of spacetime and independent gauge transformations of $A_1, A_2$. This is  ensured by  $I_{\rm hydro}$ being a local action of $h_{1,2}, \tau, B_{1,2}$.
%depends on the background fields and dynamical variables $X_{1,2}^\mu, \vp_{1,2}$ only through $h_{1,2}$ and $B_{1,2}$.
 By construction $h_{1,2}$ and $B_{1,2}$ are invariant under  %provided we also transform
%dynamical fields $X^\mu_{1,2}, \vp_{1,2}$ accordingly
 ($s=1,2$)
\bega
 \label{1diffg}
g_{s \mu \nu}' (X_s') = {\p X_s^\lam \ov \p X_s'^\mu} {\p X_s^\rho \ov \p X_s'^\nu} g_{s \lam \rho} (X_s)   , \quad
A'_{s \mu} (X_s') = {\p X_s^\lam \ov \p X_s'^\mu} A_{s \lam} (X_s), \quad
X_s'^\mu (\sig)  = f_s^\mu (X_s (\sig))  \\
A_{s \mu}' = A_{s\mu} - \p_\mu \lam_s (X_s) , \qquad \vp_s' (\sig)= \vp_s (\sig) + \lam_s (X_s (\sig))
\label{1ga}
\end{gather}
for arbitrary functions $f_1, f_2$ and $\lam_1, \lam_2$.

$I_{\rm hydro}$ further satisfies the following symmetry conditions:
\ben

%\item Spacetime diffeomorphisms

\item Invariant under spatial and time diffeomorphisms in the fluid spacetime
 \bega \label{sdiff}
 \sig^i \to  \sig'^i (\sig^i), \qquad \sig^0 \to \sig^0  \\
 \label{tdiff}
 \sig^0 \to \sig'^0= f (\sig^0, \sig^i), \qquad \sig^i \to \sig^i   \ .
 \end{gather}
These symmetry conditions define a fluid. %, say, as compared to a solid.

\item  Invariant under a diagonal spatial-dependent shift
\be\label{cshift}
\vp_r \to \vp_r - \lam (\sig^i) , \qquad \vp_a \to \vp_a
\ee
with $\vp_r = \ha (\vp_1 + \vp_2)$ and $\vp_a = \vp_1 - \vp_2$.
This condition specifies a normal fluid as compared to a superfluid.

\item  Invariant under a $Z_2$ reflection symmetry
\be\label{keyp3}
  I^*_{\rm hydro} [h_1,  B_1; h_2, B_2;  \tau] = - I_{\rm hydro} [h_2,  B_2; h_1,  B_1; \tau]
\ee
which is needed to be consistent with the behavior of~\eqref{pager1} under complex conjugation. The condition implies that $ I_{\rm hydro}$ must be complex and in particular the imaginary part of the action must be even under exchange of $1,2$ indices. For the path integral~\eqref{qft0} to be bounded we further require that
\be\label{pos}
{\rm Im} \, I_{\rm hydro} \geq 0
\ee
for any dynamical variables and external sources.

\item  Vanish  when we set all the sources and dynamical fields of the two legs to be equal, i.e.
\be \label{keyp2}
 I [h, B; h,  B;  \tau] = 0  \ .
\ee
Equations~\eqref{keyp3}--\eqref{keyp2} are all consequences of unitary time evolution.

\item Local KMS condition which can be stated as follows.
Setting the dynamical fields to ``background'' values %\footnote{Note that  setting $X^\mu_{1,2} = \sig^a \de_a^\mu$ reduces independent diffeomorphisms of $g_{1,2}$ to those of $\sig^a$ and it then makes sense to add $g_1$ and $g_2$.}
\be \label{bakv}
X^\mu_{1,2} = \sig^a \de_a^\mu , \quad \vp_{1,2} =0, \quad e^{\tau} =  \sqrt{- g_{r00}} , \quad  g_{r00} = \ha (g_{100} + g_{200}) \ .
\ee
in the action and denoting the resulting expression as $I_s [g_1, A_1; g_2, A_2]$, we then impose that
\be \label{kms1}
I_s [g_1, g_2,  A_1, A_2] = I_s [\tilde g_1, \tilde g_2,\tilde  A_1, \tilde A_2] \
\ee
where the tilde variables are defined as
\be
\begin{split}
\tilde g_{1\mu \nu} (x) = g_{1\mu \nu} (-t + i \th, - \vx ) , \qquad & \tilde A_{1 \mu} (x) = A_{1 \mu} (-t + i \th, - \vx), \cr
\tilde g_{2\mu \nu} (x) = g_{2\mu \nu} (- t - i (\beta_0 - \th), -\vx )  , \qquad &
\tilde A_{2 \mu} (x) = A_{2 \mu} (- t - i (\beta_0 - \th), -\vx)  \ .
\label{tiV}
\end{split}
\ee
for arbitrary $\th \in [0, \beta_0]$. This condition is to ensure that for $\rho_0$ given by the thermal density matrix~\eqref{pager1} satisfies a condition\footnote{As explained in detail in~\cite{CGL}, KMS condition relates $W$
to a time-reversed one. To obtain a condition on $W$ itself one needs to combine it with a time reversal symmetry. Depending on the situation one could combine it with $\sT$ or $\sP \sT$ or $\sC\sP\sT$. For our discussion of effective theory for a charged fluid
we choose $\sP \sT$ for definiteness.} obtained by combining
the Kubo-Martin-Schwinger (KMS) condition  with $\sP \sT$
\be
\label{1newfdt1}
 %{\rm KMS \;\; condition:} \quad
W [g_1 (x), A_1 (x); g_2 (x), A_2 (x)] = W [\tilde g_1 (x), \tilde A_1 (x); \tilde g_2 (x), \tilde A_2 (x)] \ .
\ee

\een
The action $I_{\rm hydro} [h_2,  B_2; h_1,  B_1; \tau]$ is  then obtained as the most general local action consistent with the above conditions.  In particular, at the level of equations of motion the local KMS condition recovers all the standard constraints of hydrodynamics from the entropy current condition and linear Onsager relations. Furthermore, it leads to new constraints from nonlinear generalizations of the Onsager relations, and non-equilibrium fluctuation-dissipation relations.

In this paper we further develop the formulation  in a number of ways:

\ben

\item We work out in detail the classical limit of $I_{\rm hydro}$. %\footnote{Note that as emphasized in~\cite{CGL}, in the classical limit the path integrals~\eqref{qft0} survive and describe classical statistical fluctuations.}
There are many simplifications in this limit. In particular, this enables a transparent formulation of the fluctuating hydrodynamics in physical spacetime in the presence of
arbitrary external fields. It also helps to clarify issues related to field redefinitions and frame choice.

%The formulation of~\cite{CGL} was mostly in the fluid spacetime. As briefly discussed there one can also formulate the
% theory  more directly in the physical spacetime. Here we will develop the formulation in the physical spacetime in more detail, especially including couplings to external fields.

\item We replace the local KMS condition by an alternative formulation, which directly acts on the dynamical fields.
To distinguish it from~\eqref{kms1} we will refer to it as dynamical KMS symmetry (or condition).
%At the classical level, this is a $Z_2$ symmetry (in the absence of background fields) directly acting on dynamical fields.
The dynamical formulation is equivalent to the previous one in the classical limit, but is more convenient to implement
and more general. It should be applicable to {\it any states $\rho_0$ in local equilibrium} rather than just thermal density matrix perturbed by external background fields. The dynamical formulation has been recently used in~\cite{GL} to prove the local second law of thermodynamics and also leads to an explicit construction of the entropy current from a Noether-like procedure.

\item We elaborate the formulation for a conformal fluid, which contains some new features. We also give the explicit action of a neutral conformal fluid to the second order in derivatives and work out the corresponding entropy current following the general construction presented in~\cite{GL}.
%  local KMS symmetry predicts a new relation among transport coefficients at second order in derivative expansion. We will focus on the bosonic part of the action. Fermions will be discussed elsewhere.

\een

The plan of the paper is as follows.  In next section to help set up the notations we review the action of~\cite{CGL} as well as implications of the local KMS condition. In Sec.~\ref{sec:classical} we discuss the classical limit.
In Sec.~\ref{sec:cov} we present the  formulation of dynamical KMS symmetry. In Sec.~\ref{sec:kms} we discuss the implications of
dynamical KMS symmetry on the action in the classical limit and work out the entropy current to first order in derivatives.
 In Sec.~\ref{sec:frd} we discuss field redefinitions and frame choices.
 In Sec.~\ref{sec:conf} we discuss formulation of fluctuating hydrodynamics for a conformal fluid, and work out the action and the entropy current for a neutral conformal fluid to second order in derivatives. We conclude with a brief discussion of the quantum regime in Sec.~\ref{sec:dis}.

%\HL{emphasize local equilibrium}

%which is needed to ensure the full path integral (including all loops) to satisfy~\eqref{1newfdt1}. It also makes it possible to derive an entropy current with a non-negative divergences. These topics will be presented elsewhere.

%\HL{NQEFT: emphasize non-equilibrium despite using thermal density matrix (local equilibrium, non-equilibrium at large distances and times), differences in classical and quantum
%case, operator in path integral}

%KMS condition and chemical potential

%Extend the local KMS condition to ghost fields, which will ensure the full path integral (including all loops) satisfy~\eqref{1kms}.

%\be \label{1kms}
%W [\phi_{1} (x), \phi_{2} (x) ] = W [ \phi_{1} (-x) , \phi_{2} (- t - i \beta_0, -\vx )  ] \ .
%\ee

\section{Action for fluctuating hydrodynamics}  \label{sec:review}

In this section to set up the notations we review the action for fluctuating hydrodynamics for a relativistic system with a $U(1)$ global symmetry. We will only be concerned with the bosonic action.

\subsection{Action in the fluid spacetime} \label{sec:2a}

To implement separate spatial and time diffeomorphisms~\eqref{sdiff}--\eqref{tdiff},  it is convenient to
decompose $h_{1,2}$ and $B_{1,2}$  into objects which have good transformation properties under them (below $s=1,2$)
\bega \label{1deco}
h_{sab} d\sig^a d\sig^b =  - b^2_s  \le( d\sig^0 -  v_{si} d\sig^i \ri)^2 + a_{sij} d\sig^i d\sig^j, \\
B_{sa} d\sig^a = \mu_s b (d\sig^0 - v_{si} d\sig^i) + \fb_{si} d\sig^i  \ .
\label{decB}
\end{gather}
To implement~\eqref{keyp3}--\eqref{keyp2} it is convenient to further introduce symmetric and anti-symmetric combinations ($r-a$ variables)
\begin{gather}  \label{1nr}
 E_r =  \ha \le( b_1 +b_2 \ri) ,\quad E_a
 =   \log \le(b_2^{-1} b_1 \ri) , \quad V_{ai} = E_r (v_{1i} - v_{2i})  , \quad  V_{ri} = \ha E_r (v_{1i}  + v_{2i}) , \\
 a_{r ij} = \ha  (a_{1ij} + a_{2ij}), \quad \chi_a
 = \ha \log \det (a_2^{-1} a_1), \quad
\Xi =  \log \le(\hat a_2^{-1} \hat a_{1}\ri), \\
\label{1na}
\mu_r = \ha (\mu_1 + \mu_2), \quad \mu_a = \mu_1 - \mu_2, \quad
  \fb_{ri} = \ha (\fb_{1i} + \fb_{2i}) , \quad \fb_{ai} = \fb_{1i} - \fb_{2i} \
\end{gather}
where $\hat a_{1,2}$ denotes the unit determinant part of $a_{1,2}$ and thus $\Xi$ is traceless.
%Note that $\mu_{1,2}$ are the local chemical potentials associated with two segments of the CTP contour while the local velocity for each physical spacetime can be obtained by
%\be
%u^\mu_s = {1 \ov b_s} {\p X^\mu_s \ov \p \sig^0} \ .
%\ee
Note that $\tau$ should be considered as a $r$-variable.
$\mu_r$ can be identified as the local chemical potential.

%The $r$-variables reduce to standard
%quantities like velocity, local temperature and local chemical potential, while $a$-variables correspond to noises.

The most general action which satisfies the symmetries listed in the Introduction can be readily constructed using~\eqref{1nr}--\eqref{1na} and their derivatives
\be  \label{intm}
I_{\rm hydro} = \int d^d \sig \, \sqrt{a_r} E_r \, \sL \
\ee
where $\sL$ is a scalar under~\eqref{sdiff}--\eqref{tdiff}, and can be written as
a double expansion in terms of the number of $a$-type fields (i.e. expanding in noises),
and the number of derivatives. More explicitly,
\be \label{aexp}
I_{\rm hydro} = I^{(1)} +  I^{(2)} + \cdots  , \qquad \sL =  \sL^{(1)} +  \sL^{(2)} + \cdots,
\ee
where $ \sL^{(m)}$ contains $m$ factors of $a$-fields. From~\eqref{keyp3},  $\sL^{(m)}$ is pure imaginary for  even $m$ and real for odd $m$. Each  $\sL^{(m)}$ can  be further expanded in
the number of derivatives
\be \label{dexp}
\sL^{(n)} = \sL^{(n,0)} + \sL^{(n,1)} + \cdots  \
\ee
with $\sL^{(n,m)}$ containing $m$ derivatives.
 To first order in derivatives the most general $ \sL^{(1,1)} $
can be written as\footnote{This is the most general form which is valid for all dimensions.
For specific dimensions one may introduce more terms using $\ep$-symbols.}
\bega \label{kee}
\sL^{(1,1)} = - f_1 E_a + f_2  \chi_a + f_3  \nu_a
-  { \eta  \ov 2}  \Xi^{ij} D_0 a_{rij}   -   \lam_{1}  V_{a}^i D_i E_r
- \lam_2 \fc^i_a \hat D_0 \fb_{ri} + \lam_{12} V^i_a \hat D_0 \fb_{ri}
\cr  + \lam_{21}  \fc_{a}^i D_i E_r +
\lam_{5}  %\al_r^{ij}
 D_i \tau V^i_{a} + \lam_{6}  %\al_r^{ij}
 D_i \mu_r V_{a}^i
+ \lam_{7} % \al_r^{ij}
 D_i \tau \fc_{a}^i + \lam_{8} D_i \mu_r \fc_{a}^i  + \cdots,
\end{gather}
where $f_{1,2,3}$ can be further expanded in derivatives as
\bea
f_1 &=&  \ep_0 + f_{11} D_0 \tau +  f_{12}  D_0 \le(\log \sqrt{\det a_r} \ri) +  f_{13}  \beta^{-1} (\sig) D_0 \hmu  %\tr (\al_r^{-1} D_0 \al_r)
+ {\rm higher \, derivatives} , \\
f_2 &=&  p_0 + f_{21} D_0 \tau -  f_{22} D_0 \le(\log \sqrt{\det a_r} \ri) + f_{23} \beta^{-1} (\sig) D_0 \hmu
+ {\rm higher \, derivatives} , \\
f_3 &=&  n_0 + f_{31} D_0 \tau +  f_{32} D_0 \le(\log \sqrt{\det a_r} \ri) - f_{33} \beta^{-1} (\sig) D_0 \hmu
+ {\rm higher \, derivatives} \ .
\label{f3}
\eea
In the above expressions indices are raised and lowered using $a_r$, all coefficients
are all real functions of $\mu_r$ and $\beta (\sig)$, and\footnote{The choice of these combinations makes the coefficients of various terms in the expressions of the stress tensor and current simpler.}
\be \label{odef}
\nu_a = \mu_a + E_a \mu_r , \quad \hmu = \mu_r \beta (\sig), \quad  \fc_{ai} = \fb_{ai} - \mu_r V_{ai} , \quad
\hat D_0 \fb_{ri} \equiv D_0 \fb_{ri} - \mu_r D_i E_r \ .
\ee
In~\eqref{kee}--\eqref{odef} we have also used various covariant derivatives. For a scalar $\phi$ under time diffeomorphism~\eqref{tdiff}, the covariant time derivative
is defined as
\be\label{tder}
D_0 \phi \equiv {1 \ov E_r} \p_0 \phi \ .
\ee
Note that $E_r$ and $V_{ri}$ do not transform as a scalar under~\eqref{tdiff}. For them one can define a combined object
\be \label{der}
D_i E_r \equiv {1 \ov E_r} \le(\p_i E_r  + \p_0 V_{ri} \ri)
\ee
which transforms under~\eqref{tdiff} as a scalar and under~\eqref{sdiff} as a vector.
The spatial covariant derivatives for a scalar $\phi$ and a vector $\phi_i$ under~\eqref{sdiff} are defined respectively as
\bega \label{des}
D_i \phi =  \p_i \phi+ v_{ri} \p_0 \phi \equiv d_i \phi, \\
D_i\phi_j = d_i\phi_j -\tilde\Gamma^k_{ij}\phi_k,
\label{dest}
\end{gather}
where $d_i \equiv \p_i + v_{ri} \p_0$ and
\be \label{omm}
\tilde\Gamma^i_{jk} \equiv \ha a^{il}_r \left(d_j a_{rkl}+ d_k a_{rjl}
-d_l a_{rjk}\right)=\Gamma^i_{jk}+\frac{1}{2}a^{il}_r \left(v_{rj}\partial_0a_{rkl} + v_{rk}\partial_0a_{rjl}
-v_{rl}\partial_0a_{rjk}\right)
\ee
with $\Gamma^i_{jk}$ the Christoffel symbol corresponding to $a_{r ij}$.

 To zeroth order in derivatives the most general $\sL^{(2,0)}$ can be written as
\bea
- i \sL^{(2,0)} &=& s_{11}E^2_a+s_{22}\chi^2_a+s_{33}\nu^2_a + 2 s_{12} E_a \chi_a
 + 2 s_{13} E_a \nu_a  \cr
&&+ 2 s_{23}\chi_a \nu_a
+ r \tr  \Xi^2 +r_{11} V_a^i V_{ai} + 2 r_{12} V_a^i\fc_{ai} + r_{22} \fc_a^i \fc_{ai}
 \label{aa0},
\eea
where again all coefficients are real and are functions of $\mu_r, \beta (\sig)$.
Given that $\sL^{(2)}$ is pure imaginary, in order for the
path integral~\eqref{qft0} to be well defined, the coefficients of~\eqref{aa0} must be such that the expression is non-negative
for any choices of dynamical variables.

In~\eqref{kee} and~\eqref{aa0} we have not imposed~\eqref{bakv}--\eqref{kms1} whose consequences will be discussed
separately in Sec.~\ref{sec:lkms}.

\subsection{Formulation in physical spacetime} \label{sec:phy}

The action~\eqref{kee}, \eqref{aa0} is formulated in the fluid spacetime. The advantage of this formulation is that
the action can be easily coupled to external sources and the symmetries of the theory are easy to implement.
A shortcoming is that connections with the dynamics in physical spacetime and the standard form of hydrodynamic
equations  are not manifest. Alternatively one can formulate the effective action in the physical spacetime. For this purpose, introduce
\be \label{1ax}
X_a^\mu = X_1^\mu (\sig) - X_2^\mu(\sig),  \quad X^\mu (\sig) = \ha (X_1^\mu (\sig) + X_2^\mu (\sig))  \ . %= X(\sig) + \ha X_a (\sig), \quad X_2 (\sig) = X(\sig) - \ha X_a(\sig) \ .
\ee
We interpret $X^\mu (\sig)$ as the motion of a fluid element in the physical spacetime (now only one copy) and $X^\mu_a$ as  statistical and quantum noises of that motion.
One can then invert $X^\mu (\sig^a)$ to obtain $\sig^a (X^\mu)$, and treat $X^\mu$ as the coordinates of the physical spacetime  and $\sig^a (X)$ as dynamical
variables. Other dynamical variables $X_a^\mu (\sig), \vp_{r,a} (\sig), \tau (\sig)$ %$X_a^\mu (x) = X_a^\mu (x (X)), \vp_{r,a} (x (X)), \tau_{r,a} (x(X))$
are now all considered as functions of $X^\mu$ through $\sig^a (X)$. To emphasize that $X^\mu$ are now simply coordinates  and not dynamical variables we will denote them as $x^\mu$.
The dynamical variables are now
$\sig^a (x), X_a^\mu (x), \vp_{r,a} (x), \tau (x)$.
%Using time reparameterization of $\sig^0$, equation~\eqref{tdiff} we can set $\sig^0 = x^0$, and thus the remaining variables
%are  $\sig^i (x), X_a^\mu (x), \vp_{r,a} (x), \tau (x)$.

The background fields then have the form
\be
g_{1 \mu \nu} \le(x + \ha X^a (x) \ri), \quad A_{1 \mu} \le(x + \ha X^a (x) \ri), \quad
g_{2 \mu \nu} \le(x - \ha X^a (x) \ri), \quad A_{2 \mu} \le(x - \ha X^a (x) \ri)
\ee
whose arguments depend on dynamical variables $X^\mu_a$, and thus should be expanded in $X^\mu_a$ when performing the
noise expansion~\eqref{aexp} in the action. Direct formulation in physical spacetime is complicated as one does not have a canonical definition of the spacetime metric. The obvious candidate   $g = \ha (g_1 + g_2)$ does not make sense
as $g_1$ and $g_2$  transform under independent diffeomorphisms. Thus one cannot just add them. Similar statement applies to $A= \ha (A_1+A_2)$.\footnote{In next section we will see these difficulties go away in the classical limit.}
Nevertheless, one could  construct the theory by inverting the action in the fluid spacetime.
For example, in the absence of background fields one finds~\eqref{kee} can be simply written as
 %One can define the velocity field $u^\mu$ and local chemical potential $\mu$ as
%\be \label{veo}
%u^\mu = {1 \ov b} {\p X^\mu \ov \p \sig^0} , \quad b = \sqrt{-\eta_{\mu\nu}  {\p X^\mu \ov \p \sig^0} {\p X^\mu \ov \p \sig^0}  }, \quad \mu = u^\mu \p_\mu \vp_r \
%\ee
%where both quantities should be understood as functions of $x^\mu$ through inverting $X^\mu (\sig)$.
%Expanding the action in $X_a^\mu, \vp_a$, one finds~\eqref{kee} and~\eqref{aa0} become
\bega \label{kee2}
 I^{(1)} = \int d^d x \,  \le[  T^{\mu \nu}  \p_\mu X_{a \nu}
+  J^\mu \p_\mu \vp_a \ri] + O(a^3)\
\end{gather}
where $T^{\mu \nu}$ and $J^\mu$ are the hydrodynamic stress tensor and $U(1)$ current
obtained from~\eqref{kee}.

%constructed from velocity field $u^\mu, \mu$, and local temperature $T(x) = T_0 e^{-\tau (x)}$.

%where
%and (below $w_{\mu} =  \p_\mu \vp_a + 2 \mu u^\rho \p_{(\mu} X_{a \rho)}$)
%\be
%\begin{split}
%\tilde I^{(2)}_0=& i \int d^dx \, \left[ r \eta^{\mu\rho}\eta^{\nu\sigma}(2 \partial_{<\mu}X_{a\nu>})(2 \partial_{<\rho} X_{a\sigma>} ) +r_{11}\Delta^{\mu\rho}  (2 u^\nu \partial_{(\mu}X_{a\nu)}) (2 u^\sigma \partial_{(\rho}X_{a\sigma)}) \right. \\
% & + r_{22} \Delta^{\mu\nu}w_\mu w_\nu  + 2 r_{12}  \Delta^{\mu\rho} (2 u^\nu\partial_{(\mu}X_{a\nu)}) w_\rho  \\
%& +  s_{11}(u^\mu \partial X_{a\mu})^2 + s_{22} (\Delta^{\mu\nu}\partial_\mu X_{a\nu})^2 + s_{33} (\p \vp_a)^2   \\
%& -2 s_{12} \Delta^{\mu\nu}\partial_\mu X_{a\nu} u^\rho\partial X_{a\rho} + 2 s_{23} ( \p \vp_a) \Delta^{\mu\nu}\partial_\mu X_{a\nu}  -  2 s_{13}  u^\mu\partial X_{a\mu} ( \p \vp_a)
 %\bigg],
%\end{split}
%\label{1qua}
%\ee

\subsection{Consequences of local KMS condition} \label{sec:lkms}

Now let us turn to the local KMS condition~\eqref{bakv}--\eqref{kms1}. We will mention its consequences and then discuss some open issues.
Applying to~\eqref{kee} and~\eqref{aa0}, we can group consequences of~\eqref{kms1} into three types (with the first equation of~\eqref{bakv}, $\sig^a = \de_\mu^a x^\mu$ where $x^\mu = (x^0, \vx)$ denotes the spacetime coordinates):

\ben

\item For time-independent $g_{1,2}$ and $A_{1,2}$, $I^{(1)}$ should have a factorized form to leading order $g_a = g_1 - g_2$ and $A_a = A_1 - A_2$, i.e.
\be \label{fac}
I^{(1)}_s = \tilde W [g_1, A_1] - \tilde W[g_2, A_2] + O(g_a^3, A_a^3)
\ee
where $\tilde W [g (\vx), A (\vx)]$ is some local functional defined on the spatial part (with coordinate $\vx$) of the spacetime and satisfies
\be
\tilde W [g (\vx), A(\vx) ] = \tilde W [g (-\vx), A(-\vx) ] \ .
\ee
Applying~\eqref{fac} to~\eqref{kee}--\eqref{f3} we find various coefficients in $\sL^{(1)}$ should satisfy
standard thermodynamic relations
\be \label{thermo}
\ep_0 + p_0 - \mu n_0 = - {\p p_0 \ov \p \tau} , \qquad
n_0 = {\p p_0 \ov \p \mu},
\ee
or
\be
\ep_0 + p_0 = - \le({\p p_0 \ov \p \tau}\ri)_{\hmu} , \qquad
n_0 = \beta {\p p_0 \ov \p \hmu},
\ee
%\be \label{1ent}
%\hat s_0 = \ep_0+ p_0 - \mu n_0
%\ee
and
\be \label{1con}
 %\lam_{5}  = \lam_{1} , \qquad \lam_{7} = - \lam_{21 } , \qquad  \lam_{6}   =\lam_{8} =0
 \lam_{5}  = \lam_{1} + \mu \lam_{12} , \qquad \lam_{7} = - \lam_{21 } - \mu \lam_2 , \qquad  \lam_{6}   =\lam_{8} =0    \ .
 \ee
Equations~\eqref{1con}  reproduce the equality-type constraints from the entropy current.
Note that $\tilde W$ may be interpreted as the partition function of the system on a stationary manifold with metric $g_{\mu \nu} (\vx)$ and external source $A_\mu (\vx)$,  and the above discussion derives the partition function prescription proposed in~\cite{Banerjee:2012iz,Jensen:2012jh}.

\item  Non-equilibrium Onsager relations
\be \label{1con1}
\sG_{i j} (x, y; \phi_i (\vx)] = \sG_{j i} (-y, -x; \phi_i (-\vx) ] ,
\ee
where $\phi_i$ collectively denotes $g_{ \mu \nu}, A_\mu$with $i$ labeling different components of both $g$ and $A$.
$\sG_{ij}$ is %a function of $x^\mu, y^\mu$ and functional of $\phi_i (\vx)$,
defined as
\be \label{fge}
\sG_{i j} (x, y; \phi_i (\vx)] = {\de^2 I_s^{(1)} \ov \de \phi_{ai} (x) \phi_{rj} (y)} \biggr|_S
\ee
where the subscript $S$ in~\eqref{fge} denotes the procedure that after taking the differentiation one should
set $g_{1\mu \nu} = g_{2\mu \nu} =g_{\mu \nu} , A_{1\mu} = A_{2\mu} = A_\mu $ with both $g_{\mu \nu} $ and $A_\mu$ time-independent. The notation $\sG (\cdots ]$ highlights that it is a function of $x^\mu,y^\mu$, but a functional of $\phi_i (\vx)$. Applying~\eqref{1con1} to $I^{(1)}$ one finds that
\be\label{11ons}
\lambda_{12} = \lambda_{21}, \qquad
- f_{13}  = f_{31}  ,\qquad
f_{23} = f_{32}, \qquad
- f_{12}  = f_{21} \ .
\ee
Note that~\eqref{11ons} are the standard constraints from linear Onsager relations. To the first derivative order of~\eqref{kee} there is no difference whether one imposes~\eqref{1con1} on the full nonlinear action~\eqref{kee} or
the linearized version (around the equilibrium). But one expects additional nonlinear constraints to start appearing at second derivative order~\cite{CGL}.

\item Non-equilibrium fluctuation-dissipation relations which relate parameters of  $\sL^{(1)}$ and $\sL^{(2)}$:
\bega  \label{1tv}
%\begin{align}
r =\frac{\eta}{2} T(\sig), \quad
r_{11}=\lambda_1 T(\sig),\quad
r_{12} %= -\frac{\lambda_{12}+\lambda_{21}}{2}  T(\sig)
= - \lam_{12} T(\sig),\quad
r_{22}=\lambda_{2} T(\sig) , \\
%\ee
%and
%\begin{align}
\label{1sc}
 s_{11}  = {f_{11} T(\sig)} , \qquad
  s_{12} = { f_{12}  T(\sig)}  , \qquad
   s_{13} =  { f_{13}  T(\sig)}, \qquad
\\
s_{22}
=  f_{22}T(\sig), \qquad
s_{23}
=-\frac{f_{32}+f_{23}}{2} T(\sig) = - { f_{23}T(\sig)},\qquad
 s_{33}
=f_{33} T(\sig) ,
\label{2sc}
%\end{align}
\end{gather}
where $T(\sig)$ was introduced in~\eqref{tay0}.
When applying to a system near equilibrium, i.e. setting $T(\sig)$ and $\mu (\sig)$ to equilibrium values equations~\eqref{1tv}--\eqref{2sc} reproduce precisely the standard  fluctuation-dissipation relations, yet here they are derived for arbitrary  $\tau (\sig^a)$ and $\mu (\sig^a)$ and thus are valid for far-from-equilibrium situations\footnote{Recall all the coefficients
are functions of $\tau (\sig^a)$ and $\mu (\sig^a)$.}.
As discussed below~\eqref{aa0}, the coefficients of~\eqref{aa0} have to be such that right hand side of~\eqref{aa0}
is always non-negative. From relations~\eqref{1tv}--\eqref{2sc}, this translates into the statement
that dissipative parameters such as the conductivity, shear and bulk viscosities are non-negative~\cite{CGL}.
This reproduces the inequality-type constraints from the entropy current.

\een
Note that to the orders of derivative expansion given in~\eqref{kee} and~\eqref{aa0} there is no difference between quantum
or classical regime. Neither are the results sensitive to $\theta_0$.

The local KMS prescription~\eqref{bakv}--\eqref{kms1} works very well, it reproduces of all the known
constraints on hydrodynamical equations, predicts new nonlinear Onsager relations as well as non-equilibrium fluctuation-dissipation relations.
%It has also been checked in~\cite{CGL} that
 %they indeed ensure KMS conditions~\eqref{1newfdt1} for all two-point functions.
 Yet there are
still a number of  deficiencies. Firstly there is an ambiguity in its formulation. A key element of~\eqref{bakv} is to set the ``background value'' of
$e^\tau$ to $\sqrt{-g_{r00}}$ motivated from that in a curved spacetime with metric $g_{\mu \nu}$ the local proper temperature
is proportional to ${1 \ov \sqrt{-g_{00}}}$. The value $\sqrt{- g_{r00}}$ was chosen  for $e^\tau$ as it is symmetric in $1,2$ and reduces to $\sqrt{-g_{00}}$ when $g_1 = g_2 = g$.
The choice of is clearly not unique. For example, another possibility is\footnote{Note that at the level of~\eqref{kee} and~\eqref{aa0} using~\eqref{abd} does not make a difference.}
\be \label{abd}
 \tau = {1 \ov 4} \le( \log (- g_{100}) + \log (- g_{200}) \ri) \ .
\ee
Secondly, while there have been many supporting evidences that the prescription~\eqref{bakv}--\eqref{kms1} indeed ensures the KMS condition~\eqref{1newfdt1}, there has not been a general proof. Thirdly,  the condition~\eqref{1newfdt1} should be physically equivalent regardless of choice of $\th$ in~\eqref{tiV}. But in the
the formulation~\eqref{kms1} this is not clear. %\HL{Note that to the orders of derivative expansion given in~\eqref{kee} and~\eqref{aa0}  the results are insensitive to $\theta$. }
Finally,~\eqref{kms1} is not formulated directly in the dynamical variables
%\footnote{The formulation relies on the feature of $I_{\rm hydro}$ that
%couplings of $I_s$ obtained from~\eqref{bakv} determine the couplings of the full action $I_{\rm hydro}$. \HL{This may not be true for example in the presence of anomalies.} },
which makes its implementation and use inconvenient.
In particular it relies on the feature that in $I_{\rm hydro}$ the dynamical variables and background fields always come together in the form of $h_{1,2}$ and $B_{1,2}$. This feature will likely not hold in the presence of anomalies.

In Sec.~\ref{sec:classical} we discuss the classical limit of~\eqref{qft0} and show that the first three issues mentioned in the previous paragraph are all
addressed in that limit. In Sec.~\ref{sec:cov} we introduce a dynamical formulation  which acts directly on the dynamical variables.

\section{Classical limit}  \label{sec:classical}

The path integrals~\eqref{qft0} describe macroscopic behavior of quantum systems with a nonzero $\hbar$.
%When the temperature is sufficiently high where quantum fluctuations are small, we can consider the classical limit $\hbar \to 0$.
In this section we consider the classical limit $\hbar \to 0$.
Note in the  classical limit  the path integrals~\eqref{qft0} survive and describe classical statistical fluctuations.
%As discussed in~\cite{CGL}, unlike in the usual situation where the $\hbar \to 0$ limit leads to the saddle point of the path integral (i.e. equations of motion),
%the path integrals~\eqref{path} and~\eqref{qft0} survive the $\hbar \to 0$ limit with an effective $\hbar_{\rm eff}$ controlled by
%the energy or entropy density of the system. That is, now the path integrals integrate over all possible thermal statistical fluctuations. In this section we work out the general structure of the action~\eqref{intm} in the $\hbar \to 0$ limit.  We will find great simplifications for the physical spacetime formulation.

\subsection{Small $\hbar$ expansion}

Following the discussion of~\cite{CGL}, reinstating $\hbar$ we can write various background and dynamical fields as
\bega \label{re1}
g_{1\mu\nu}=g_{\mu\nu}+\frac  \hbar2 g_{a\mu\nu} ,\quad g_{2\mu\nu} =g_{\mu\nu}-\frac \hbar2  g_{a\mu\nu}, \quad
A_{1\mu}=A_{\mu}+\frac \hbar 2 A_{a\mu} , \quad A_{2\mu}=A_{\mu}-\frac \hbar 2  A_{a\mu} \\
X_1^\mu =X^\mu+\frac \hbar2   X_a^\mu ,\quad  X_2^\mu =X^\mu-\frac \hbar 2  X_a^\mu ,\quad
\varphi_1 =\varphi+\frac \hbar 2   \varphi_a ,\quad  \varphi_2=\varphi-\frac \hbar 2 \varphi_a ,
\label{re2}
\end{gather}
and $\th, \beta_0$ in~\eqref{tiV} become $\hbar \th, \hbar \beta_0$.  Furthermore, suppose $f^{(n)}$ is a coefficient in $n$-th order action $I^{(n)}$ in the $a$-field (noises) expansion~\eqref{aexp}, then $f^{(n)}$ can be expanded in $\hbar$ as
\be
f^{(n)} = {1 \ov \hbar^{n-1}} \le(f^{(n)}_{\rm cl} + O(\hbar)  +  \cdots \ri) \ .
\ee
In~\eqref{1diffg}--\eqref{1ga}, various transformation parameters can be written as
\begin{align}
f_1^\mu=f^\mu+\frac 12 \hbar  f_a^\mu ,\qquad  f_2^\mu=f^\mu-\frac 12 \hbar f_a^\mu , \qquad \lam_1=\lam+\frac 12 \hbar  \lam_a , \qquad  \lam_2 =\lam-\frac 12 \hbar \lam_a\ .
\end{align}

In the  $\hbar \to 0$ limit, the two diffeomorphisms~\eqref{1diffg} then become: (i)  physical space diffeomorphisms
\be\label{ddeo}
X^\mu \to X'^\mu (X) = f^\mu (X), %\quad X_a'^\mu (X') = \p_\lam f^\mu X_a^\lam (X)
\ee
under which $X^\mu_a$ transform as a vector, $g_{\mu \nu}, g_{a \mu \nu}$ as symmetric tensors, and $A_\mu, A_{a \mu}$ as one-forms, and (ii) noise diffeomorphisms under which various quantities transform as
\be \label{upo}
X_a'^\mu (\sig) = X_a^\mu (\sig) + f_a^\mu (X (\sig)), \quad g_{a \mu \nu}'   = g_{a \mu \nu} - \sL_{f_a} g_{\mu \nu} , \quad
A_{a \mu }'    = A_{a \mu}  - \sL_{f_a} A_\mu  \ .
\ee
where $\sL_w$ denotes Lie derivative along a vector $w^\mu$.
We emphasize that~\eqref{upo} are finite transformations.
The gauge transformations~\eqref{1ga} become physical spacetime gauge transformation
\be \label{upo1}
A_{\mu}'   =  A_{\mu}  - \p_{\mu} \lam (X) , \quad  A_{a \mu}' (X)  =  A_{a\mu} (X)  + \p_\mu \le(\sL_{X_a} \lam \ri) ,\quad \vp' (\sig) = \vp (\sig) + \lam (X(\sig)) ,
\ee
and noise gauge transformation
\be
A_{a \mu}' (X)  =  A_{a\mu} (X)  - \p_{\mu} \lam_a (X) , \qquad \vp_a' (\sig) = \vp_a (\sig)
+ \lam_a (X(\sig)) \ .
\label{upo3}
\ee
We then find that in this limit
\be \begin{split}
h_{1ab}& =\p_a X_1^\mu\p_b X_1^\nu g_{1\mu\nu}(X_1)
= h_{ab} (\sig) +  {\hbar \ov 2} h^{(a)}_{ab} + O(\hbar^2) ,  \\
B_{1a}&=\p_a X_1^\mu A_{1\mu}(X_1) + \p_a \vp_1 = B_a (\sig)+ {\hbar \ov 2} B^{(a)}_a + O(\hbar^2) ,
\label{b1}
\end{split}
\ee
where
\bega\label{pp1}
h_{ab} (\sig) \equiv  \p_a X^\mu\p_b X^\nu g_{\mu \nu} (X) , \qquad h^{(a)}_{ab} = \p_a X^\mu\p_b X^\nu G_{a \mu \nu} (X), \\
B_a  \equiv  \p_a X^\mu A_\mu (X) + \p_a \vp (\sig), \qquad B^{(a)}_a  = \p_a X^\mu C_{a \mu}  (X) \\
G_{a\mu\nu} (X) %= \nabla_\mu X_{a\nu}+\nabla_\nu X_{a\mu}+g_{a\mu\nu}(X)
 \equiv g_{a\mu \nu} + \sL_{X_a} g_{\mu \nu}
, \qquad
C_{a\mu} %= A_{a\mu}(x)+ \p_\mu \vp_a (x) +  \p_\mu X_a^\nu A_\nu(x)+  \p_\nu A_\mu(x)X_a^\nu
\equiv  A_{a\mu}(X )+ \p_\mu \vp_a (X) + \sL_{X_a} A_\mu \ ,
\label{pp2}
\end{gather}
and $\vp_a (X) \equiv \vp_a (\sig (X))$.  %$\sL_{X_a}$ denotes the Lie derivative along the vector $X_a^\mu (\sig (X))$.
It can be readily checked that $C_{a\mu}$ and $G_{a \mu \nu}$ are invariant under~\eqref{upo}--\eqref{upo3} and transform
as a vector and tensor respectively under~\eqref{ddeo}. The corresponding equations for $h_2,  B_2$ are obtained from~\eqref{b1} by switching the signs before the $O(\hbar)$ terms.  We now have
\be
{1 \ov \hbar} I_{\rm hydro} [h_1, B_1; h_2, B_2; \tau]  = I_{\rm hydro} [h_{ab}, B_a; h^{(a)}_{ab}, B^{(a)}_a; \tau] + O(\hbar) \ .
\ee

%the $r$-type variables are kept to $O(\hbar^0)$ while the $a$-type variables are kept to order $O(\hbar)$.
As before decompose $h_{ab}$ and $B_a$ as
\bega \label{ddeco}
h_{ab} d\sig^a d\sig^b =  - b^2  \le( d\sig^0 -  v_{i} d\sig^i \ri)^2 + a_{ij} d\sig^i d\sig^j, \\
B_{a} d\sig^a = \mu b (d\sig^0 - v_{i} d\sig^i) + \fb_{i} d\sig^i  \ .
\end{gather}
and
\begin{gather}  \label{tbe0}
{\p X^\mu \ov \p \sig^0} \equiv b u^\mu, \quad u^\mu u_{ \mu} = -1, \quad
u_{ \mu} = g_{\mu \nu}  u^\nu ,\quad
{\p X^\mu \ov \p \sig^i} \equiv - v_i  b u^\mu + \lam_i{^\mu}, \quad u_\mu \lam_i{^\mu} = 0 \
\end{gather}
where $u^\mu$ is the local velocity field.
We find various quantities in~\eqref{1nr}--\eqref{1na} become
\begin{gather}
E_r = b, \qquad a_{r ij} = a_{ij} , \qquad V_{ri} = V_i = b v_i  \qquad  \fb_{ri} =  \fb_i, \qquad \mu_r = \mu \\
 E_a  =-\ha   u^\mu u^\nu G_{a \mu \nu} (X)
 , \quad V_{ai} = u^\mu \lam_i^\nu G_{a \mu \nu}   , \quad
 \Xi  = \le(\lam^{i \mu} \lam_j^\nu -{\De^{\mu \nu}  \ov d-1} \de_i^j \ri) G_{a \mu \nu}   , \\
 \chi_a
  = \ha \De^{\mu \nu} G_{a \mu \nu}
\quad \mu_a = u^\mu C_{a \mu} + \ha \mu u^\mu u^\nu G_{a \mu \nu} , \quad
 \quad \fb_{ai} = \lam_i^\mu C_{a \mu} + \mu u^\mu \lam_i^\nu G_{a \mu \nu}
\end{gather}
and
\be \label{odef1}
\nu_a = \mu_a + E_a \mu_r = u^\mu C_{a \mu}  ,  \qquad  \fc_{ai} = \fb_i - \mu_r V_{ai} = \lam_i^\mu C_{a \mu} \ .
\ee
As anticipated from~\eqref{pp1}--\eqref{pp2} all $a$-type fields can be obtained from $C_{a\mu}$ and $G_{a \mu \nu}$ .

%For $r$-type variables we have kept to zeroth order in $\hbar$ and for $a$-type variables we have scaled a factor of $\hbar$.

Now let us consider the KMS condition~\eqref{1newfdt1}--\eqref{tiV} in the  $\hbar \to 0$ limit. Equations~\eqref{tiV} can be written as
\bega \label{ckms1}
\tilde \phi (x)  = \phi (-x) , \qquad \tilde \phi_a (x) = \phi_a (-x) +i \sL_{\beta_0} \phi (-x), \qquad \beta_0^\mu = \beta_0 \le({\p \ov \p x^0} \ri)^\mu
\end{gather}
where $\phi = \{ g_{\mu \nu}, A_\mu \}$ and $\phi_a = \{ g_{a\mu \nu}, A_{a\mu}\}$.
Note that parameter $\th$ has dropped out. Equation~\eqref{1newfdt1} can be written as
\be \label{kmsc1}
W [\phi (x), \phi_a (x)] = W [\phi (-x), \phi_a (-x) +i \sL_{\beta_0} \phi (-x) ] \ .
\ee
We emphasize that in~\eqref{kmsc1} the shift in $\phi_a$ is a finite transformation so one cannot expand the right hand side
in $\sL_{\beta_0} \phi (-x) $. Note that under~\eqref{bakv}
we have
\be \label{pi3}
h_{ab} = g_{\mu \nu} \de_a^\mu \de_b^\nu, \quad h^{(a)}_{ab} = g_{a \mu \nu} \de_a^\mu \de_b^\nu, \quad
 B_a  = A_\mu \de_\mu^a , \quad B^{(a)}_a = A_{a \mu} \de_\mu^a
\ee
thus the local KMS prescription~\eqref{kms1} implies that the action satisfies
\be \label{oo}
I_{\rm hydro} [h_{ab}, B_a; h^{(a)}_{ab}, B^{(a)}_a; \tau]  = I_{\rm hydro} [\tilde h_{ab}, \tilde B_a; \tilde h^{(a)}_{ab}, \tilde B^{(a)}_a; \tilde \tau]
\ee
where
\bega\label{pi5}
\tilde h_{ab} (\sig) = h_{ab} (-\sig), \quad \tilde B_a (\sig) = B_a (-\sig), \quad \tilde \tau (\sig) = \tau (-\sig)  \\
  \tilde h^{(a)}_{ab} (\sig) =  h^{(a)}_{ab} (-\sig) + i \beta_0 \p_0 h_{ab} (-\sig), \quad
  \tilde B^{(a)}_a (\sig) =  B^{(a)}_a  (-\sig) + i \beta_0 \p_0 B_{a} (-\sig) \ .
  \label{pi4}
  \end{gather}
From an argument given in Appendix~\ref{app:pro} we can then immediately conclude from~\eqref{oo} that the KMS condition~\eqref{kmsc1} is satisfied at tree level.

%Note that $\vp_1, \vp_2$ are defined in fluid spacetime, i.e. they are defined to be scalar functions of $\sig^a$.
%Now with $\sig^a$ treated as scalar functions of $x^\mu$ they become scalar functions of $x^\mu$. They do not transform
%under~\eqref{diff2}. Similarly with $\tau$ which becomes a scalar function of $x^\mu$.
%Note that had $\vp_{1,2}$ been defined as scalar fields in each physical spacetimes, then $\vp_a$ will also transform
%under~\eqref{diff2} as discussed in Appendix~\ref{app:a}.

\subsection{Physical space formulation}

In the $\hbar \to 0$ limit the physical spacetime formulation is much simplified.
In fact the fluid and physical spacetime formulations become essentially the same. In this subsection, rather than starting from the fluid spacetime formulation we present an intrinsic formulation for the fluid action in the physical spacetime itself.

In the classical limit, in the physical spacetime  the dynamical variables are $\sig^a (x), \vp (x), \tau (x)$ and $ X^\mu_a (x), \vp_a (x)$. The background fields are $g_{\mu \nu} (x),  A_\mu(x), g_{a \mu \nu} (x), A_{a \mu} (x)$ with $g_{\mu \nu}$ the physical spacetime metric.
The action should be invariant under: (i)  physical spacetime diffeomorphism~\eqref{ddeo}; (ii) noise diffeomorphism~\eqref{upo}; (iii)  gauge transformation~\eqref{upo1}; (iv) noise gauge transformation~\eqref{upo3}; (v) time and spatial diffeomorphisms of $\sig^a$~\eqref{sdiff}--\eqref{tdiff} which are now ``global'' symmetries; (vi) the diagonal shift~\eqref{cshift} which is also now a global symmetry;  (vii) equations~\eqref{keyp3} and~\eqref{keyp2} which are now imposed on physical spacetime action; (viii) the local KMS condition.

(ii)-(iv) imply that $a$-fields (including both background and dynamical variables) must appear in the combinations $G_{a\mu \nu}, \; C_{a \mu}$ introduced in~\eqref{pp2}, as these are the only combinations invariant under~\eqref{upo}--\eqref{upo3}, while $A_\mu$ and $\vp$ must appear through  $B_\mu = A_\mu + \p_\mu \vp (x)$. By using the time diffeomorphism~\eqref{tdiff} we can set $\sig^0 = x^0$. In the absence of parity or time reversal breaking, invariance under~\eqref{sdiff}  implies that
the only invariant which can be constructed is the velocity field
$u^\mu$  defined by
\be \label{1vel}
u^\mu = {1 \ov \sqrt{-j^2}} j^\mu, \quad j^2 \equiv j^\mu j_\mu , \quad
j^\mu = \ep^{\mu \mu_1 \cdots \mu_{d-1}} {\p \sig^{1} \ov \p x^{\mu_1}} \cdots {\p \sig^{d-1} \ov \p x^{\mu_{d-1}}}
\ee
where $\ep$ is the antisymmetric {\it tensor} and indices are raised and lowered by $g_{\mu \nu}$ and its inverse.
Note that $j^2$ is not invariant under~\eqref{sdiff} and by definition
\be
u^\mu u_\mu = -1  \ .
\ee
It can be readily checked the definition~\eqref{1vel} coincides with that in~\eqref{tbe0}. $B_\mu$ is not invariant under shift~\eqref{cshift} of $\vp$, but
\be
\mu \equiv u^\mu B_\mu, \qquad F_{\mu \nu} = \p_\mu B_\nu - \p_\nu B_{\mu}
\ee
are invariant. Note that $F_{\mu \nu}$ does not depend on the dynamical variables.

To summarize,  the only combinations of $r$-variables which can appear are
\be \label{ovar}
\beta (x) = \beta_0 e^{\tau (x)}, \quad u^\mu, \quad \mu , \quad  F_{\mu \nu} , \quad g_{\mu \nu} \ .
\ee
Sometimes it is convenient to combine the first three variables further into
\be\label{uen}
\beta^\mu = \beta (x) u^\mu (x),  \qquad  \hat \mu (x) = \beta (x) \mu = \beta^\mu (x)  B_\mu
\ee
where  $\beta^\mu$ is now unconstrained. Any scalar functions in the action must only depend on $\mu$ and $\beta (x)$.
%\be
%- \beta^\mu \beta_\mu =  \beta^2 (x) \ .
%\ee

Now introducing  notation
\be \label{n0}
G_{a\mu M}=(G_{a\mu\nu},2 C_{a\mu}), \qquad M = (\mu, d), \quad G_{a \mu d} = 2 C_{a \mu}
\ee
we can write the full action as
\be
I_{\rm hydro} = \int d^d x \, \sqrt{-g} \, \sL
\ee
with %($\eta_n =0$ for $n$ odd and $\eta_n =1$ for $n$ even)
\be \label{w3}
\sL = \sum_{n=1}^\infty \sL^{(n)} = \sum_{n=1}^\infty  i^{\eta_n} f^{(n)} [\Lam_r] G_a^n  , \qquad
\eta_n = \bca 1 & n\; {\rm even} \cr
             0 & n \; {\rm odd}
             \eca \
%f^{(1)} [\Lam_r] G_a +i  f^{(2)} [\Lam_r] G_a^2 +  f^{(3)} [\Lam_r]G_a^3 + \cdots + i f^{(2n)} [\Lam_r] G_a^{2n} + f^{(2n+1)} [\Lam_r] G_a^{2n+1} + \cdots
\ee
where $\Lam_r$ denotes the collection $\Lam_r = \{\beta^\mu, \hat \mu, F_{\mu \nu}, g_{\mu \nu}\}$ and
we have suppressed all spacetime indices. The $n$-th term in~\eqref{w3} should be understood as
\be \label{uips}
f^{(n)} [\Lam_r] G_a^n  =  f^{(n)}_{\mu_1 \cdots \mu_n, M_1, \cdots M_N}
(\Lam_r; \p_\mu) \, G_{a \mu_1 M_1} (x)  \cdots G_{a \mu_n M_n} (x)\
\ee
where  the notation $  f^{(n)}_{\mu_1 \cdots \mu_n, M_1, \cdots M_N}
(\Lam_r; \p_\mu) $ indicate it is a function of $\Lam_r$, their derivatives, as well as derivative operators acting on $G_{a \mu M}$.
The whole action should be diffeomorphism invariant.
The first few terms can be written explicitly as
\be \label{acttja}
\mathcal L= \ha T^{\mu M}G_{a\mu M}+ {i \ov 4} W^{\mu\nu,MN} G_{a\mu M}G_{a\nu N}+ {1 \ov 8} Y^{\mu\nu\rho,MNP}G_{a\mu M}G_{a\nu N}G_{a\rho P}+\cdots \ ,
\ee
with
\be \label{newf}
\ha T^{\mu M}G_{a\mu M} =  T^{\mu \nu} \le(\ha g_{a \mu \nu} + \nab_\mu X_{a \nu} \ri) + J^\mu \le(A_{\mu a} + \p_\mu \vp_a +
X_a^\nu \nab_\nu A_\mu + A_\nu \nab_\mu X_a^\nu \ri)  \ .
\ee
In the first term of~\eqref{acttja} by integrating by part  we can move all the derivatives on $G_{a \mu M}$ into $T^{\mu M} = (T^{\mu \nu}, J^\mu)$.  One should keep in mind that $W, Y$ still contain derivatives on $G$'s.
From coupling to $g_{a \mu \nu}$ and $A_{a \mu}$ we can thus identify
$T^{\mu \nu}$ and $J^\mu$  as the hydrodynamic stress tensor and $U(1)$ current.

Note that in derivative counting, $u^\mu, \mu, \beta , G_{a \mu M}$ should all be counted as zeroth order.

In the absence of $g_{a \mu \nu}$ and $A_{a\mu}$, the equations of motion of~\eqref{w3} by varying with respect to $r$-variables can be consistently solved by setting $X_a^\mu = \vp_a =0$. The nontrivial equations of motion arise from varying with respect to $X_a^\mu$ and $\vp_a$, and only the first term in~\eqref{acttja} is relevant leading to
\be
\label{eom}
%E^\mu \equiv
 \nab_\nu T^{\mu \nu} - F^{\mu \nu} J_\nu = 0 , \qquad % E_d \equiv
  \nab_\mu J^\mu = 0 \ .
\ee

The action~\eqref{acttja} can also be reached by starting from the fluid spacetime action and inverting $X^\mu (\sig)$. In particular,  applying discussion parallel to that of Appendix F of~\cite{CGL} one can prove that all coefficients in~\eqref{acttja} can indeed be expressed  only in terms of $\beta^\mu, \hat \mu, F_{\mu \nu}, g_{\mu \nu}$.
The implications from the local KMS condition discussed earlier in Sec.~\ref{sec:lkms} for the fluid spacetime action can also be translated to~\eqref{acttja}.  In Sec.~\ref{sec:kms} we will present an alternative way to work out those constraints
using the new dynamical formulation introduced in next section.

Now let us  comment on the relation with previous literature. $G_{a \mu \nu}$ and $C_{a \mu}$ already appeared in~\cite{Harder:2015nxa}\footnote{$G_{a\mu\nu}$ is equal to $\xi_{\mu\nu}^a$, which is defined at the beginning of Sec. 4.2, and $C_{a\mu}$ is $\chi_\mu^a$, which is defined below eq. (4.16) there.} as well as the $O(a)$ part of the action~\eqref{acttja}. But in~\cite{Harder:2015nxa} it was not clear how to extend the action beyond $O(a)$ at nonlinear level. Here we show that these quantities are in fact exact. Also local KMS condition was not discussed there.

The physical spacetime formulation (in the gauge $\sig^0 = x^0$) shares some common elements with  the formulation of~\cite{Dubovsky:2005xd,Dubovsky:2011sj,Endlich:2012vt,Dubovsky:2011sk,Endlich:2010hf,Nicolis:2011ey,Nicolis:2011cs,Delacretaz:2014jka,Geracie:2014iva,Grozdanov:2013dba}, in addition to having noise fields $X^\mu_a, \vp_a$, the key differences are that: (i) in~\eqref{sdiff} we require general spatial diffeomorphisms while in~\cite{Dubovsky:2005xd,Dubovsky:2011sj,Endlich:2012vt,Dubovsky:2011sk,Endlich:2010hf,Nicolis:2011ey,Nicolis:2011cs,Delacretaz:2014jka,Geracie:2014iva,Grozdanov:2013dba} only volume-preserving diffeomoprhisms are allowed; (ii) we have an additional scalar field $\tau$ which serves as local temperature. Suppose we only require volume-preserving
diffeomoprhisms in~\eqref{sdiff}, then $\sqrt{-j^2}$ in~\eqref{1vel} is invariant and becomes a dynamical variable which naively may be used to replace $\tau$. However, by definition $j^\mu$ is  exactly conserved regardless of the presence of dissipation. Such a conserved quantity appears to have no place in dissipative hydrodynamics.
In our construction the full spatial diffeomorphisms get rid of $j^2$, and we supplement that by introducing $\tau$. The resulting $\beta^\mu$ is then unconstrained.

%\HL{Check that $u^\mu$ is a vector under spacetime diffeomorphisms and work out how $u^\mu$ transform under parity and time inversions.}

% \HL{Are $\mu, F$ independent?}

%\HL{Also check our previous $O(a^2)$ action is indeed the most general.}

\section{Dynamical KMS symmetry}  \label{sec:cov}

 In this section we introduce an alternative to the local KMS condition. The new formulation, to which we refer as dynamical KMS condition (or symmetry), directly acts on dynamical variables.
We first discuss the proposal at finite $\hbar$ in the fluid spacetime and then discuss the classical limit.
At the end we discuss some open issues at  quantum level.
 %It is formulated at finite $\hbar$, but there are various potential ambiguities. We then discuss the classical limit, which is
 %clean.

 %We first discuss its formulation in the fluid spacetime and then the formulation in physical spacetime.
%Compared to the formulation of last section, this version is covariant with respect to the symmetries of the Lagrangian and easier to work with. In particular it enables a simple proof that the path integral satisfies the KMS condition at tree-level.

 \subsection{Proposal}

We propose that in the absence of background fields, the action $I_{\rm hydro}$ is invariant under  the following  transformations on  the dynamical variables  ($s=1,2$)
\be \label{12}
\tilde X^\mu_s (- \sig) = - \Phi_{i \eta_s}^* X^\mu_s ( \sig) - i \eta_s \de^\mu_0, \qquad
\tilde \vp_s (-\sig) = - \Phi_{i \eta_s}^* \vp_s ( \sig), \qquad \tilde \tau (\sig) =  \tau (-\sig)\
\ee
with $\eta_1 = - \th, \, \eta_2 = \beta_0 - \th$.
In~\eqref{12}, $\Phi_{\lam}$ is a one-parameter ($\lam$)  diffeomorphism generated by vector field
\be \label{v2}
w^a = {e^\tau \ov b_r}  \le({\p \ov \p \sig^0} \ri)^a   , \qquad b_r = \sqrt{- h_{00}}
\ee
and $\Phi^*_\lam$ denotes its push-forward map.  To implement~\eqref{12} one needs to analytically continue
$\Phi^*_\lam$ to complex values of $\lam$. %Also note that the notation $\Phi_{\lam}^* X^\mu (- \sig)$ should be understood as
%first evaluating $\Phi_{\lam}^* X^\mu (\sig)$ and then take $\sig^a \to - \sig^a$.
In~\eqref{12} the constant shifts in $X^\mu_s$ are chosen so that in the presence of background fields,  if we transform the background fields as \eqref{tiV}, $h_{sab}$ and $B_{sa}$ transform as
\bega
 \label{iikms}
\tilde h_{sab} (\sig) = \Phi_{i \eta_s}^*  h_{sab} (- \sig) , \qquad \tilde B_{sa} (\sig) = \Phi_{i \eta_s}^*  B_{sa} (-\sig) , \qquad
%\tilde \tau (\sig) = \tau (-\sig)
%\\
%  \label{iikms1}
 %\tilde B_{2a} (\sig) = \Phi_{i \beta_0}^* B_{2a} (-\sig) , \qquad \tilde h_{2ab} (\sig) =  \Phi_{i \beta_0}^*
% h_{2ab} (-\sig)
\end{gather}
and the fluid spacetime action is invariant
\be \label{lkms}
I_{\rm hydro} [h_1,  B_1; h_2,  B_2, \tau] = I_{\rm hydro}  [\tilde h_1,  \tilde B_1; \tilde h_2, \tilde B_2 ; \tilde \tau] \ .
\ee
%When the external fields $g_{1,2}, A_{1,2}$ are turned on, transformation~\eqref{12} is no longer a symmetry
%of the action (as background fields do not transform under a symmetry), but~\eqref{iikms}--\eqref{lkms} still apply if one supplements~\eqref{12} by transformation~\eqref{tiV} of the background fields.
Now using a general result in Appendix~\ref{app:pro} we immediately conclude that~\eqref{12} ensures~\eqref{1newfdt1} at tree-level of the path integral~\eqref{qft0}. To ensure~\eqref{1newfdt1} at the level
of the full path integral (i.e. including all loops), one needs to extend transformations~\eqref{12} to fermionic fields, which will be discussed elsewhere.

In~\eqref{v2} the factor $1/b_r$ is inserted  so that $w^a$ is independent of choice of $\sig^0$.
 $\Phi_\lam$ is a time diffeomorphism which can be written as
\be
u^0_\lam \equiv u^0 (\sig^a; \lam), \qquad u^i = \sig^i
\ee
where $u^0 (\sig^a; \lam)$ is obtained by solving the differential equation
\be \label{d1}
{d u^0 \ov d\lam} =  {e^\tau \ov b_r} (u^a)  , \qquad u^0 (\lam=0) = \sig^0  \ .
\ee
We then have
\be \label{d2}
\Phi_{\lam}^* X^\mu_s (\sig) = X^\mu_s  (u^a_{-\lam} (\sig)) , \qquad %= u^a_{-\lam} (\sig) \de_a^\mu + \pi^\mu (u^a_{-\lam} (\sig))
\Phi_{\lam}^* \vp_s (\sig) = \vp_s  (u^a_{-\lam} (\sig))
\ .
\ee

 At a heuristic level, one may interpret the action of $\Phi^*_{i \eta_s }$ as
shifting $\hat \sig^0$ by $i \eta_s e^\tau$ where $\hat \sig^0$ is a proper time defined by $d \hat \sig^0 = b_r d \sig^0$.%\footnote{
%This is heuristic as such a $\hat \sig^0$ does not exist (since the equation
%$d \hat \sig^0 = b_r d \sig^0$ is in general not integrable) and $e^\tau$ is a function of $\sig^a$.}
We stress that, as in~\eqref{bakv}, it is the appearance of $e^\tau$ in~\eqref{v2} that ``defines'' it  as the local inverse temperature.
In other words, if we had used some other function of $\tau$ in~\eqref{v2} then it would be that function which should be identified as the local temperature.

We should mention that if we replace $\Phi^*_{i \eta_s}$ in~\eqref{12} by $\Phi^*_{\lam_s}$ with some arbitrary parameter $\lam_s$, then one will still get~\eqref{iikms} with again $i \eta_s$ replaced by $\lam_s$. Furthermore one can still use the argument of Appendix~\ref{app:pro} to conclude that~\eqref{12} ensures~\eqref{1newfdt1} at tree-level of the path integral~\eqref{qft0}.
We now show that $\lam_s$ are in fact required to be $i \eta_s$
for $X^\mu_s$ to have the right boundary conditions. We require at spacetime infinities both the background and dynamical fields go to zero, i.e. at spacetime infinities, the physical and fluid spacetime coincide and  the system is in thermal equilibrium.
Write
\be
X^\mu (\sig) = \sig^a \de_a^\mu + \pi^\mu (\sig)
\ee
then $\pi^\mu$ (which does not have to be small) should go to zero at the spacetime infinities of the fluid spacetime.
In~\eqref{d1} as $u^a \to \infty$, we should have $w^0 \to 1$ and thus
\be\label{d3}
u^0_\lam (\sig \to \infty) = \sig^0 + \lam , \qquad u^i = \sig^i  \ .
\ee
Now using~\eqref{d2} and~\eqref{d3} we thus find that if we use $\Phi_{\lam_s}$ in~\eqref{12}
\be \label{15}
\tilde X^\mu_s (\sig \to \infty) = - u_{-\lam_s}^a (- \sig) \de_a^\mu - i \eta_s \de^\mu_0 =  \sig^a \de_a^\mu + (\lam_s - i \eta_s) \de^\mu_0, \qquad \sig \to \infty \ .
\ee
We thus conclude
\be
\lam_s = i \eta_s \ .
\ee

\subsection{The classical limit}

Let us look at transformations~\eqref{12} in the classical limit $\hbar \to 0$.
%In the process as a consistency check we will also derive~\eqref{iikms} from~\eqref{11}--\eqref{12}.
With the notations introduced in Sec.~\ref{sec:classical} we find\footnote{We note that despite some resemblance of eqs. (\ref{x1}) and (\ref{pi1}) with the $U(1)_T$ transformation of \cite{Haehl:2015uoc} (see e.g. eq. (5.2)  there), they are fundamentally different.
Here due to an additional spacetime reflection, the transformation is a discrete $Z_2$ transformation. Many consequences of the theory depend crucially on this feature. In contrast a $U(1)$ transformation will lead to completely different (physically inconsistent) results.}
\bega \label{x1}
\tilde X^\mu (\sig) = - X^\mu (-\sig), \qquad \tilde X^\mu_a (\sig) = - X_a^0 (-\sig)
-i  \beta^\mu (-\sig)+ i \beta_0^\mu  \\
%\tilde X^i (\sig) = - X^i (-\sig), \qquad \tilde X^i_a (\sig) = - X_a^i (-\sig)
%- {i \beta (- \sig)  \ov b (-\sig)} \p_0  X^i (-\sig) \\
\tilde \vp  (\sig) = - \vp (-\sig), \qquad \tilde \vp_a (\sig) = - \vp_a (-\sig)-
 i \beta^a \p_a  \vp (-\sig) \
 \label{x2}
\end{gather}
where again $\th$ has dropped out and
\be
\beta (\sig) = \beta_0 e^{\tau (\sig)}, \qquad \beta^a \equiv \beta_0 w^a , \qquad \beta^\mu = \beta (\sig) u^\mu (\sig)
= \p_a X^\mu \beta^a \ .
\ee
We then find that
\bega \label{pi1}
\tilde h_{ab} (\sig) =  h_{ab} (-\sig) , \quad
\tilde h^{(a)}_{ab} (\sig) = h^{(a)}_{ab} (- \sig) + i  \sL_\beta h_{ab} (-\sig) \\
 \tilde B_{a} (\sig) = B_{a} (-\sig) , \qquad \tilde B^{(a)}_{a} (\sig) = B^{(a)}_{a} (- \sig) + i  \sL_\beta B_{a} (-\sig) \
 \label{pi2}
\end{gather}
where $\sL_\beta$ is the Lie derivative along vector $\beta^a$.

Note that one can use the time diffeomorphism~\eqref{tdiff} to set
\be \label{ppp}
\sqrt{-h_{00}} = e^\tau
\ee
in which case
\be
\beta^a = \beta_0 \le({\p \ov \p \sig^0}\ri)^a
\ee
and then the Lie derivatives in~\eqref{pi1}--\eqref{pi2} become ordinary derivatives along the time direction in the fluid spacetime.
Equation~\eqref{ppp} is a constraint between local temperature and $X^\mu (\sig)$ which makes them no longer independent.

One can readily check that the above transformation is $Z_2$.
The above equations can be written more explicitly in terms of
various components as
\be
\tilde \Phi (\sig) = \Phi (- \sig), \quad {\rm for} \quad \Phi = \{b, v_i , a_{ij}, \mu, \fb_i\}  \ ,
\ee
and
\bega
\tilde E_a (-\sig) = E_a (\sig) +  i \beta (\sig) D_0  \tau (\sig) , \quad \tilde \chi_a (-\sig) = \chi_a (\sig) + { i } \beta (\sig) D_0 \log \sqrt{a} (\sig)\\
\tilde \Xi (-\sig) = \tilde \Xi (\sig)+ i \beta (\sig)  \le(a^{-1} D_0 a \ri)_{\rm traceless} (\sig) , \quad
\tilde \fb_{ai}  (-\sig) =   \fb_{ai} (\sig) + i \beta (\sig) D_0  \fb_{i} (\sig), \\
\tilde \mu_{a}  (-\sig) =   \mu_{a} (\sig) + i \beta (\sig)  D_0  \mu(\sig)  , \quad
\tilde V_{ai} (-\sig) = V_{ai} (\sig)  +  i \beta (\sig)  \le(D_i E (\sig)   -   D_i  \tau  (\sig)  \ri)
 \\
%\label{hum} \\
\tilde \nu_a (-\sig) = \nu_a ( \sig) + i  D_0  \hmu (\sig) , \quad
\fc_{a i} (-\sig) = \fc_{ai} (\sig) + i (\beta \hat D_0 \fb_i +\hmu D_i \tau)  \ .
\end{gather}

Note that with $e^\tau = \sqrt{-g_{r00}}$, the vector field~\eqref{v2} is given by
\be
w^0 =  1 + O(\hbar)
\ee
and equations~\eqref{pi1}--\eqref{pi2} become~\eqref{pi5}--\eqref{pi4}. We thus conclude that
in the classical limit, the new prescription is precisely equivalent to~\eqref{bakv}--\eqref{kms1}.
As a cross check, applying~\eqref{lkms} with~\eqref{pi1}--\eqref{pi2} to~\eqref{kee} and~\eqref{aa0}
we find indeed identical results to~\eqref{thermo}--\eqref{2sc}. %\HL{See Appendix~\ref{app:check} for more details.}

Finally let us write down the dynamical KMS transformation for fields in physical spacetime.
From~\eqref{x1}--\eqref{x2} we immediately have
\bega \label{si1}
\tilde \sig^i (x) = - \sig^i (-x) , \quad \tilde \tau (x) = \tau (-x) , \quad \tilde \vp  (x) = - \vp (-x), % \quad
  \\
\tilde X^\mu_a (-x) = - X^\mu_a (x) - i \beta^\mu (x) + i \beta_0^\mu  , \qquad
\tilde \vp_a (-x) = - \vp_a (x)-
 i \beta^\mu \p_\mu \vp  (x) \
\label{si3}
\end{gather}
and thus
\bega \label{si4}
\tilde u^\mu (x) = u^\mu (-x), \qquad \tilde \hmu (x) = \hmu (-x),
\qquad
\tilde \beta^\mu (x) = \beta^\mu (-x),  \\
\label{si2}
\tilde G_{a\mu\nu} (-x) =  G_{a\mu \nu}  (x) + \sL_{i \beta^\mu} g_{\mu \nu} (x) ,  \qquad
\tilde C_{a\mu} (-x) =  C_{a \mu} (x) + \sL_{i \beta^\mu} B_\mu  (x)\ .
\end{gather}
The above transformations are again $Z_2$. Using the unified notations of~\eqref{n0} we can also write~\eqref{si2} as
\be \label{glkms}
\tilde G_{a \mu M} (-x) = G_{a\mu M}  (x) + \sL_{i \beta^\mu} g_{\mu M} (x)=
G_{a\mu M}  (x) + i \Phi_{r \mu M}
\ee
where $g_{\mu M} = (g_{\mu \nu}, 2 B_\mu)$ and
\be
\Phi_{r \mu \nu} = \nab_{\mu} \beta_{\nu} + \nab_{\nu} \beta_{\mu}, \qquad \Phi_{r \mu d} = 2 \sL_{ \beta^\mu} B_\mu  (x) =
2 \le(\nab_\mu \hmu - \beta^\nu F_{\mu \nu}  \ri)\ .
\ee

\subsection{Open issues at finite $\hbar$}

As in the prescription for $e^\tau$ in~\eqref{bakv}, here the choice of $1/b_r$ factor in~\eqref{v2} is not unique.
Other than that we need something to make $w^a$  independent of choice of $\sig^0$, we do not have
other criterion to fix it further at the moment. For example, instead of using $b_r$ we could have used $\sqrt{b_1 b_2}$ or even
use $b_1$ for $X_1$ and $b_2$ for $X_2$. In the classical limit all these choices become the same and there is a unique $b$.
In the classical limit $\th$ drops out and the dynamical KMS transformation is a $Z_2$ symmetry of the action.
At  finite $\hbar$, under~\eqref{v2} it does not appear that~\eqref{12} is a $Z_2$ transformation, and it is also not clear whether different $\th$ yields the same physics.
% (they should). %In Appendix~\ref{app:z2} we examine the issues of $Z_2$ and $\th$-equivalence in some detail.

\section{Dynamical KMS invariance and entropy current} \label{sec:kms}

To elucidate further the structure of the action for fluctuating hydrodynamics at classical level, in this section we
work out the implications of invariance of~\eqref{w3}  under dynamical KMS transformations~\eqref{si4}--\eqref{si2}.
We will also explicitly construct the entropy current to first derivative order using the procedure of~\cite{GL}.
%and \HL{clarify issues related to field redefinitions and frame choices. }

\subsection{Dynamical KMS invariance}

Under a dynamical KMS transformation~\eqref{si4}--\eqref{glkms} we find~\eqref{w3} becomes %(below $\eta_n = 1$ for $n$ even and $0$ for $n$ odd)
\be \label{sl0}
\tilde \sL =  % f^{(1) *} [\Lam_r] (G_a + i \Phi_r)   + i  f^{(2)*} [\Lam_r] (G_a + i \Phi_r) ^2  + \cdots +
\sum_{n=1}^\infty \widetilde{\sL^{(n)}} = \sum_{n=1}^\infty  i^{\eta_n}  f^{(n)*} [\Lam_r] (G_a + i\Phi_r)^n %+ \cdots
\ee
where $f^{(n)*}$ is obtained from $f^{(n)}$ with a sign flip on derivatives
\be
f^{(n) *}_{\al_1 \cdots \al_n} (\Lam_{r} (x); \p_\mu) =  f^{(n)}_{\al_1 \cdots \al_n} (\Lam_{r} (x); - \p_\mu)  \ .
\ee
The dynamical KMS condition can then be written as
\be \label{dkms0}
\sL = \tilde \sL - \nab_\mu V^\mu \ .
\ee
Taking another tilde operation on the above equation and from its $Z_2$ nature we find that
\be\label{inVV}
\tilde V^\mu = V^\mu  \
\ee
where the tilde operation on $V^\mu$ should understood  as in~\eqref{sl0}, i.e. one replaces $G_a$ by
$G_a + i\Phi_r$ and then flips the sign of all the derivatives. Note that this is slightly different from the tilde operation~\eqref{si1}--\eqref{glkms} defined for individual fields. Below throughout the paper tilde operations on a quantity which is part of a Lagrangian should always be understood this way.

We can expand $V^\mu$ in terms of the number of $a$-fields and derivatives
\be \label{vv}
V^\mu = \sum_{n=0}^\infty V_n^\mu = \sum_{n,m=0}^\infty  i^{\eta_n} V^\mu_{(n,m)} % + V^\mu_1 +  \cdots + i^{\eta_n} V^\mu_n + \cdots
\ee
where $V^\mu_n$ contains $n$ factors of $G_a$, and $V^\mu_{(n,m)}$ contains $n$ factors of $G_a$ and $m$ derivatives.
Equation~\eqref{dkms0} can then be written order by order in $a$-expansion as
\be \label{nex}
\sL^{(n)} = \le(\tilde \sL\ri)_n - \p_\mu V_{n}^\mu  \
\ee
where $(\tilde \sL)_n$ denotes $O(a^n)$ terms in $\tilde \sL$. In particular, for $n=0$ we have
\be \label{nnm}
\le(\tilde{\cal L} \ri)_0 = \p_\mu V^\mu_{0} \ .
\ee

Since $\Phi_{r }$ contains one derivative, the dynamical KMS condition~\eqref{dkms0} couple $n$-th derivative terms in $f^{(1)}$ with $(n-1)$-th derivative terms in $f^{(2)}$, $(n-2)$-th derivative terms in $f^{(3)}$, etc. all the way to zeroth derivative terms in $f^{(n)}$. More explicitly, using the notation of~\eqref{dexp}, $\sL^{(n,m)}$ with a fixed $l= n + m$ couple to one another.   Introducing
\be
\sL = \sum_{n=1}^\infty\sum_{m=0}^\infty \sL^{(n,m)} = \sum_{l=1}^\infty \sL_l, \quad \sL_l = \sum_{n+m=l} \sL^{(n,m)}, \quad
V^\mu = \sum_{l=0}^\infty v^\mu_l , \quad v^\mu_l  = \sum_{n+m=l} i^{\eta_n} V^\mu_{(n,m)}
\ee
we then find that~\eqref{dkms0} reduces to $\sL_l$ being separately invariant
\be \label{moon}
\tilde \sL_l -  \sL_l = \p_\mu v^\mu_{l-1} , \qquad l =1,2, \cdots \ .
\ee
When one considers a truncation of $\sL$ in derivative and $a$-expansion one should do it in terms of $\sL_l$ to be compatible with the dynamical KMS symmetry.

%For example, at order $O(a^0)$ we find
%\bega \label{gkm1}
% f^{(1) *} [\Lam_r]   \Phi_r -  f^{(2)*} [\Lam_r] \Phi_r^2 + \cdots
% + (-1)^{[{n\ov 2}]} f^{(n)*} [\Lam_r] \Phi_r^n + \cdots = \nab_\mu V^\mu_0
 % \end{gather}
%where $[x]$ denotes the integer part of $x$. It can be shown that $V_n$ for $n \geq 1$ can be set to zero by adding total derivative terms to $\sL$. This is not possible for $V_0$ in~\eqref{gkm1} as $\sL$ does not contain any term at $O(a^0)$.
%At order $O(a)$ as discussed below~\eqref{?acttja} we will need to perform a further integration by parts so that there is no derivative on $G_a$  which can generate a nonzero $V_1^\mu$.

There is a simple way to impose dynamical KMS invariance~\eqref{nex}
at order $O(a^n)$ with $n \geq 1$, which also shows  that one can set $V_n^\mu$ with $n \geq 1$ to zero by absorbing such total derivatives into the definition of the Lagrangian\footnote{See Appendix~\ref{app:totd} for an alternative argument}.  For this purpose let us consider a Lagrangian density $\mathcal L_c$ of the form (\ref{w3}), then due to $Z_2$ nature of the dynamical KMS transformations,
\be\label{w4}
\mathcal L=\frac 12\left(\mathcal L_c+\tilde{\mathcal L}_c\right),
\ee
where $\tilde{\mathcal L}_c$ is defined as in~\eqref{sl0}, automatically satisfies dynamical KMS invariance. Note, however, that $\tilde{\mathcal L}_c$ contains terms with $r$-fields only, and the resulting $\mathcal L$ violates the condition (\ref{keyp2}). We must then further require that $O(a^0)$ terms in $\tilde{\mathcal L}_c$ be equal to zero,  which is simply~\eqref{nnm}. Thus the combination of~\eqref{w4} and~\eqref{nnm} is enough to ensure~\eqref{dkms0}.

With a $\sL$ built from~\eqref{w4},  only $V_0^\mu$ is nonzero. In particular, from~\eqref{inVV} $V_0^\mu$ should contain only  even derivative terms as odd derivative terms change sign under tilde operation at $O(a^0)$. Furthermore, one can show that the even derivative part of~\eqref{nnm} is automatically implied by~\eqref{w4} and thus one  needs to consider only the odd derivative part of~\eqref{nnm}.  Now using~\eqref{moon} we can write~\eqref{nnm} more explicitly as
\be \label{gg0}
\le(\tilde \sL_{2n+1}\ri)_0 = \le(\widetilde{\sL^{(1,2n)}} + \widetilde{\sL^{(2,2n-1)}} + \cdots
+ \widetilde{\sL^{(2n+1,0)}} \ri)_0 = i \nab_\mu V^\mu_{(0,2n)} \ , \quad n=1, 2, \cdots  \
\ee
or in terms of notations of~\eqref{w3}
\bega \label{gkm1}
\sum_{k=1}^{2n+1} \ep_k f^{(k,2n+1-k) } [\Lam_r]   \Phi_r^k %-  f^{(2,2n-1)*} [\Lam_r] \Phi_r^2 + \cdots
 %+ (-1)^{n} f^{(2n+1,0)*} [\Lam_r] \Phi_r^n + \cdots
 =  i \nab_\mu V^\mu_{(0,2n)}
  \end{gather}
where $\ep_k = (-1)^m$ for $k=2m+1, 2m+2$ and $f^{(k,l)}$ denotes terms with $l$ derivatives.

After imposing~\eqref{w4} and~\eqref{nnm} we will need to perform a further integration by parts for $f^{(1)}$ terms in the Lagrangian~\eqref{w3} so that in the first term of~\eqref{acttja} there is no derivative acting on $G$.
This can generate a nonzero $V_1^\mu$.

As an illustration of the procedures outlined above, let us consider~\eqref{acttja} up to
\be
\sL = \sL_1 + \sL_2 + \sL_3 + \cdots
\ee
i.e.  to two derivatives in $T^{\mu M}$, one derivative in $W^{\mu \nu, MN}$, and zero derivative in $Y^{\mu \nu \rho, MNP}$. We will use the notation $ T^{\mu M}_n$ to denote terms in $ T^{\mu M}$ which contain $n$ derivatives, and similarly for others.
We thus have
\bega \label{iio}
\sL_1 = \ha T_0^{\mu M} G_{a \mu M} , \qquad \sL_2= \ha T^{\mu M}_1 G_{a\mu M}+ {i \ov 4} W^{\mu\nu,MN}_0 G_{a\mu M}G_{a\nu N} , \\
\mathcal L_3= \ha T^{\mu M}_2 G_{a\mu M}+ {i \ov 4} W^{\mu\nu,MN}_1 G_{a\mu M}G_{a\nu N}+ {1 \ov 8} Y^{\mu\nu\rho,MNP}_0 G_{a\mu M}G_{a\nu N}G_{a\rho P} \ .
\label{iii}
\end{gather}

Applying~\eqref{w4} we find
\bega \label{uio0}
T^{\mu M}_1  = - \ha W^{\mu \nu, MN}_0 \Phi_{r \mu N} , \qquad  W^{\mu  \nu, M N}_1 = {3 \ov 4}  Y^{\mu  \nu  \rho, MN P}_0  \Phi_{r\rho P}
 \ .
\end{gather}
We next turn to $O(a^0)$ dynamical KMS condition~\eqref{gkm1} which can be written explicitly as
\be \label{002}
\ha T^{\mu M *} \Phi_{r \mu M} -  {1 \ov 4} W^{\mu \nu, MN * }  \Phi_{r \mu M}   \Phi_{r \nu N}
- {1 \ov 8}  Y^{\mu \nu \rho, MNP *} \Phi_{r \mu M}   \Phi_{r \nu N}     \Phi_{r \rho P}   + \cdots = \nab_\mu V^\mu_0  \ .
\ee
At first order derivative order~\eqref{002} gives
\be \label{003}
\ha T^{\mu M}_0  \Phi_{r \mu M} = \nab_\mu V^\mu_{(0,0)} \
\ee
where the second subscript of $V$ denotes the number of derivatives. At second derivative order~\eqref{002} is automatically satisfied from~\eqref{uio0} with $ V^\mu_{(0,1)} =0$, and at third derivative order using~\eqref{uio0} we have
\be \label{004}
\ha T^{\mu M }_2 \Phi_{r \mu M} + {1 \ov 16}  Y^{\mu \nu \rho, MNP}_0 \Phi_{r \mu M}   \Phi_{r \nu N}     \Phi_{r \rho P}
= \nab_\mu V^\mu_{(0,2)}  \ .
\ee
Note that the above expression also indicates that to second order in $T^{\mu M}$ there is no derivative acting on $G_{a \mu M}$
in~\eqref{acttja}. So there is no  need to do further integration-by-parts
and to the current order $V_1^\mu =0$.

In~\cite{GL} we showed for any theory of the form~\eqref{w3} (in the absence of $a$-type sources $g_{a \mu \nu}$ and $A_{a \mu}$)  which satisfies~\eqref{pos} and is invariant under dynamical KMS transformations~\eqref{si4}--\eqref{si2}, there exists a current $S^\mu$ whose divergence is non-negative.  From~\cite{GL}, to second order in derivative expansion $S^\mu$ can be written as
\be \label{s0}
S^\mu = V_{(0,0)}^\mu + V_{(0,2)}^\mu  -T^{\mu\nu}\beta_\nu-J^\mu\hat\mu  \ .
\ee
Below we will work out its explicit form to first derivative order for a general charged fluid and show it indeed reproduces the standard entropy current. In next section we will work out its explicit expression at second derivative order for a conformal neutral fluid.

\subsection{Explicit tensor analysis to first order in derivative expansion}

We now expand various tensors above explicitly in terms of $\tau (x), u^\mu , \hmu, g_{\mu \nu}, F_{\mu \nu}$.
The analysis becomes rather tedious at second order in derivatives for $T^{\mu M}$. So we will only write down the explicit
expressions for $T^{\mu M}$ to first order in derivatives. In Sec.~\ref{sec:conf} we give the explicit expression at second derivative orders for a conformal neutral fluid.

To first derivative order the most general $T^{\mu M}  = (T^{\mu \nu}, J^\mu)$ can be written as
 \be\label{1gens}
T^{\mu \nu} = \ep u^\mu u^\nu + p \De^{\mu \nu} +
2  u^{(\mu} q^{\nu)} + \Sig^{\mu \nu} ,
\qquad J^\mu = n u^\mu + j^\mu,
\ee
with
\bega
 \label{1fird}
\ep = \ep_0 + h_\ep , \quad p= p_0 + h_p, \quad \Sig^{\mu \nu} = - \eta \sig^{\mu \nu}, \quad
n = n_0 + h_n,   \quad
\De^{\mu \nu} = \eta^{\mu \nu} + u^\mu u^\nu \\
 \label{1p}
 h_\epsilon =
f_{11} \p\tau +  f_{12}  \th +  f_{13} \beta^{-1} (x) \p \hmu , \\ h_p =  f_{21} \p \tau - f_{22} \th +  f_{23} \beta^{-1} (x) \p \hmu  , \\
h_n  = f_{31} \p\tau + f_{32}   \th -   f_{33} \beta^{-1} (x)   \p \hmu  \\
 j^\mu = \lam_{21} \p u^\mu
 -\lam_2  \le( \De^{\mu \nu} \p_\nu  \mu + u_\lam  F^{\lam \mu} \ri)+ \lam_7 \De^{\mu \nu} \p_\nu  \tau +\lam_{8} \De^{\mu \nu} \p_\nu  \mu \\
  \label{11p}
 q^\mu = - \lam_{1}  \p u^\mu
  +\lam_{12}  \le( \De^{\mu \nu} \p_\nu  \mu + u_\lam  F^{\lam \mu} \ri)+ \lam_5 \De^{\mu \nu} \p_\nu  \tau +\lam_{6} \De^{\mu \nu} \p_\nu  \mu , \\
 \p \equiv u^\mu \nab_\mu, \quad \th\equiv \nab_\mu u^\mu, \quad \sig^{\mu \nu} \equiv \De^{\mu \lam} \De^{\nu \rho} \le(\nab_\lam u_\rho + \nab_\rho u_\lam - {2 \ov d-1} g_{\lam \rho}
\nab_\al u^\al \ri) \
\label{qcur4}
\end{gather}
where all coefficients are functions of $\tau$ and $\hmu$. We have used notations to coincide with the stress tensor and current following from~\eqref{kee}.

 At zeroth derivative order, equation~\eqref{003} requires $\ep_0, p_0, n_0$ satisfy the standard thermodynamic relations
\be \label{thermo1}
\ep_0 + p_0  = - {\p p_0 \ov \p \tau} , \qquad
n_0 = \beta  {\p p_0 \ov \p \hmu}   ,
\ee
with
\be\label{entr1}
V_{(0,0)}^\mu= p_0 \beta^\mu \ .
\ee
In other words, equation~\eqref{003} imposes local first law of thermodynamics.

To examine implications of the first equation of~\eqref{uio0}  we also need to write down the most general form of
\be \label{yip}
W_0^{\mu \nu, MN} = W_0^{\nu \mu, NM}
\ee
with zero derivative.  More explicitly we can then write
\begin{align}
\label{W01}
W_0^{\mu\alpha,\nu\beta}&= s_{11}u^\mu u^\nu u^\alpha u^\beta+ s_{22}\Delta^{\mu\nu}\Delta^{\alpha\beta} - s_{12}(u^{\mu} u^{\nu}\Delta^{\alpha\beta}+u^{\alpha} u^{\beta}\Delta^{\mu\nu}) \cr
&\ \ + 2 r_{11}\left(u^\mu u^{(\alpha}\Delta^{\beta)\nu}+u^\nu u^{(\alpha}\Delta^{\beta)\mu}\right)+ 4 r
\le(\Delta^{\alpha(\mu}\Delta^{\nu)\beta} - {1 \ov d-1}\Delta^{\mu\nu}\Delta^{\alpha\beta}  \ri)\\
W_0^{\mu\alpha,\nu d}&=-  s_{13}u^\mu u^\nu u^\alpha+   s_{23}\Delta^{\mu\nu} u^\alpha+ 2 r_{12}
u^{(\mu}\Delta^{\nu)\alpha} , \quad  W_0^{\mu\alpha,d \nu} =  W_0^{\alpha \mu,\nu d} \\
\label{W02}W_0^{\mu\nu,dd}&= s_{33}u^\mu u^\nu+ r_{22}\Delta^{\mu\nu} \ .
\end{align}
Again we have chosen notations to be consistent with fluid spacetime action~\eqref{aa0}.
It is also convenient to decompose  $\Phi_{r\mu M}$
in terms of transverse traceless tensors, transverse vectors, and scalars with respect to $u^\mu$
\bega \label{u1}
\le(\De_\mu{^\lam} \De_\nu{^\rho} - { \De_{\mu \nu}  \ov d-1}\De^{\lam \rho } \ri)  \Phi_{r \mu \nu}
=  \beta  \sig_{\mu \nu}, \quad \De_\mu{^\nu} u^\lam   \Phi_{r \nu \lam}  = \beta v_{1 \mu} , \quad
u^\mu u^\nu \Phi_{r \mu \nu} = - 2 \beta  \p \tau \\
\De^{\mu \nu} \Phi_{r \mu \nu} = 2 \beta  \th, \quad \De_\mu{^\nu}  \Phi_{r \nu d}  = 2 \beta v_{2 \mu}, \quad
u^\mu \Phi_{r \mu d} =2  \p \hmu
\label{u2}
\end{gather}
where
\bega \label{9de}
 v_{1\mu} = \p u_\mu - \De_\mu{^\nu} \p_\nu \tau, \qquad v_{2 \mu} =  \beta^{-1} \De_\mu{^\nu} \nab_\nu \hmu - u^\nu F_{\mu \nu} \ .
 %\quad \phi_1 = - \p \tau, \quad \phi_2 = \th, \quad \phi_3 = \p \hmu
\end{gather}
Now plugging~\eqref{W01}--\eqref{9de} into the first equation of~\eqref{uio0} and comparing with the first order part of~\eqref{1gens}, we again recover~\eqref{1con},~\eqref{11ons}, and~\eqref{1tv}--\eqref{2sc}.
Note that the Onsager relations follow from~\eqref{yip}.

\subsection{Entropy current}

Using~\eqref{entr1}, to first order in derivative expansion the entropy current~\eqref{s0} has the form
\be
S^\mu
=p_0\beta^\mu-T^{\mu\nu}\beta_\nu-J^\mu\hat\mu
\ee
which recovers the standard result.
Taking the divergence of the above expression, using equations of motion~\eqref{eom} and~\eqref{uio0},~\eqref{003}
we find that
\be
\nab_\mu S^\mu =  {1 \ov 4} W_0^{\mu\nu,MN} \Phi_{r \mu M} \Phi_{r \nu N} \equiv Q_2
\ee
which agrees with the general result of~\cite{GL}. $Q_2 \geq 0$ follows from~\eqref{pos}. Now using~\eqref{W01}--\eqref{9de} we can write the right hand side of the above equation as
\bega
Q_2 =  \beta^2 \left[ r \sig^{\mu \nu} \sig_{\mu \nu}+r_{11} v_1^2 + r_{22} v_2^2 + {2} r_{12} v_1 \cdot v_2 \ri. \cr % v_1^\mu v_{2 \mu}
 \le.
  +  s_{11} (\p \tau)^2  + s_{22} \th^2 + {s_{33} \ov \beta^2} (\p \hmu)^2
 + 2 s_{12} \th \p \tau +  {2 s_{23}  \ov \beta} \th \p \hmu +   {2 s_{13} \ov \beta} \p \tau \p \hmu
 \ri] \ .
\end{gather}
Using the ideal fluid equations of motion
\be
v_{1 \mu} = - {n_0 \ov \ep_0 + p_0} v_{2 \mu}, \qquad \p\tau = \left(\frac{\p p_0}{\p\varepsilon_0}\right)_{n_0} \theta
 , \qquad {1 \ov \beta} \p\hmu = - \left(\frac{\p p_0}{\p n_0}\right)_{\varepsilon_0} \theta
\ee
in $Q_2$ then we find
\be
\label{entp}
\nabla_\mu S^\mu =
\frac \beta {2}\eta \sigma^{\mu\nu}\sigma_{\mu\nu}+\beta \zeta \theta^2+\sigma \beta  v_2^2 \ ,
\ee
where
\bega
\label{sio1}
\sigma=\frac \beta {(\varepsilon_0+p_0)^2}\left(r_{11}n_0^2-2 r_{12}n_0 (\ep_0 + p_0) +r_{22}(\ep_0 + p_0)^2\right),\\
\zeta=\beta \left(s_{11}(\p_\varepsilon p_0)^2+s_{22}+s_{33}(\p_n p_0)^2 +2s_{12}\p_\varepsilon p_0-2 s_{13}\p_\varepsilon p_0\p_n p_0-2 s_{23}\p_n p_0 \right),
\label{ze1}
\end{gather}
with $\p_\varepsilon p_0 \equiv \left(\frac{\p p_0}{\p \varepsilon_0}\right)_{n_0}$ and $\p_n p_0 \equiv \left(\frac{\p p_0}{\p n_0}\right)_{\varepsilon_0}$. In the first term of~\eqref{entp} we have used the first equation of~\eqref{1tv}, and $\sigma, \zeta$ are precisely the
expressions for conductivity and bulk viscosity identified in~\cite{CGL}, see equations (5.113) and (5.114) there.\footnote{To compare~\eqref{ze1} with (5.115) of~\cite{CGL} note that $M_{1,2,3}$ defined there are related to $\p_\varepsilon p_0$ and $\p_n p_0$ as
 \be \left(\frac{\p p_0}{\p \varepsilon_0}\right)_{n_0}=-\frac{M_1}{M_2},\quad\left(\frac{\p p_0}{\p n_0}\right)_{\varepsilon_0}=-\frac{M_3}{M_2}\ee.} We have thus recovered the standard form for the divergence of the entropy current.

\section{Frame choices from field redefinitions} \label{sec:frd}

Going to higher orders in derivative expansion the analysis becomes very tedious.
Even at the order of $T_1^{\mu M}$ and $W_0^{\mu \nu, MN}$, the Lagrangian is already pretty long.
In this section we show that the Lagrangian can be greatly simplified by taking advantage of field redefinitions.

\subsection{General discussion of field redefinitions}

Let us write the Lagrangian in the form
\be \label{yup}
\mathcal L=\sL_1+\sL_r
\ee
where  $\sL_1$ given by~\eqref{iio},
%\be
%\sL_0 = \frac 12 T_0^{\mu\nu} G_{a\mu\nu}+J_0^\mu C_{a\mu}, \qquad
% \sL_1 = \frac 12 T_1^{\mu M} G_{a\mu M}  + {i \ov 4} W_0^{\mu \nu, MN} G_{a \mu M} G_{a \nu N}
%\ee
$T_0^{\mu\nu}=\varepsilon_0 u^\mu u^\nu+p_0\Delta^{\mu\nu}$ and $J_0^\mu=n_0 u^\mu$ are the ideal stress tensor and current, and $\sL_r$ denotes the rest of the Lagrangian.  Note that the separation in~\eqref{yup} is natural as  $\sL_1$ and $\sL_r$ are invariant separately under the dynamical KMS condition, i.e.
\be \label{dkm1}
\tilde \sL_1 - \sL_1 = i \nab_\mu V_{(0,0)}^\mu, \qquad \tilde \sL_r - \sL_r = \p_\mu V^\mu_r, \qquad V_r^\mu = V^\mu - i V^\mu_{(0,0)}
\ee
We will denote the equations of motion of $\sL_1$ as
\be
E_\mu =0, \qquad E_1 =0, \qquad E_2 =0 ,
\ee
where
\be \label{0eom}
 E_\mu \equiv (\ep_0 + p_0)  v_{1 \mu}  + n_0 v_{2 \mu} , \quad  E_1 \equiv - \p \tau + \left(\frac{\p p_0}{\p \varepsilon_0}\right)_{n_0} \th ,\quad
E_2 \equiv {1 \ov \beta} \p \hmu + \left(\frac{\p p_0}{\p n_0}\right)_{\vep_0} \th \ .
\ee
Note that $E_\mu$ is the transverse part of the first equation of~\eqref{eom} while $E_{1,2}$ are related to the longitudinal part of the first equation and the second equation by a linear transform.\footnote{More explicitly, $\nabla_\mu J^\mu =-\p_\tau n_0 E_1+\beta\p_{\hat\mu}n_0 E_2$ and $u_\nu \nabla_\mu T^{\mu\nu} =\p_\tau\epsilon_0 E_1-\beta\p_{\hat\mu}\epsilon_0 E_2$.}

Let us first consider field redefinitions of the $a$-fields
\be\label{re1}
X_a^\mu\to X_a^\mu+\delta X_a^\mu, \qquad \varphi_a\to\varphi_a+\delta\varphi_a \ ,
\ee
under which
\be
G_{a \mu \nu} \to G_{a \mu \nu} + 2 \nab_{(\mu} \de X_{a \nu)}  , \quad
G_{a \mu d} \to G_{a \mu d} + 2 \p_\mu \de \vp_a + 2 \le(\p_\mu (\de X^\nu_a A_\nu) - F_{\mu\nu} \de X^\nu_a \ri)  \ .
\ee
 $\delta X_a^\mu$ and $\delta\varphi_a$ can be expanded in terms of number of derivatives and $a$-fields, e.g.
 \be
 \de X_{a \mu} = \de X^{(1)}_{a\mu} + i \de X^{(2)}_{a \mu} + \de X^{(3)}_{a \mu} + \cdots
 \ee
 where $\de X^{(n)}_{a \mu}$ contains $n$ factors of $G_{a \mu M}$ and all terms should start at zero derivative order. Similarly with $\de \vp_a$.
 The corresponding $\de G^{(n)}_{a \mu M}$ have similar expansions and all terms start at first derivative order.
 Under~\eqref{re1}, $\sL_1$ is invariant and
\bega \label{iop}
\sL_r \to   \sL_r + \de_a \sL_1 , \quad
\de_a \sL_1 =  \nab_\mu N^\mu_{a} - E_\mu \de X^\mu_a  - E_1 \de \lam_{a1} - E_2 \de\lam_{a2}
\equiv  \nab_\mu N^\mu_{a}  - E_\al \de X^\al_a \\
\label{iop1} E_\al \equiv (E_\mu, E_1, E_2), \qquad \de X^\al_a \equiv (\de X^\mu_a, \de \lam_{a1}, \de \lam_{a2})
\end{gather}
where  $\de_a \sL_r$ has been reabsorbed into $\sL_r$,
$N^\mu_{a} = T^{\mu \nu}_0 \de X^\mu_a  +J^\mu_0 \le(\de \vp_a + \de X^\nu_a A_\nu\ri)$, $\lam_{a1} , \lam_{a2}$ are
linear combinations of $u_\mu \de X^\mu_a$ and $\de \vp_a + \de X^\nu_a A_\nu$,\footnote{More explicitly,
\be
\delta\lambda_{a1}=-\p_\tau \varepsilon_0 u^\alpha\delta X_{a\alpha}-\p_\tau n_0(\delta\varphi_a+\delta X_a^\nu A_\nu),\quad \delta \lambda_{a2}=\beta\p_{\hat \mu}\varepsilon_0 u^\alpha\delta X_{a\alpha}+\beta\p_{\hat\mu}n_0(\delta\varphi_a+\delta X_a^\nu A_\nu) \ .
\ee}
and $E_\mu, E_{1,2}$ were defined in~\eqref{0eom}.  In Sec. (\ref{sec:lan}) we will show how one can use~\eqref{iop} to set to zero terms in $\sL_r$
proportional to zeroth order equations of motion~\eqref{0eom}, or proportional to their derivatives.

Let us now turn to field redefinitions of $r$-fields.
Since the full Lagrangian depends on the $r$-type dynamical fields only through $u^\mu,\ \beta (x) $ and $\mu$, we
can in fact consider nonlocal changes of $r$-variables $\sig^i, \vp$ as far as the corresponding changes
in $u^\mu, \mu$ are local. Consider
field transformations $\sigma^i\to \sigma^i+\delta \sigma^i$, $\tau\to \tau+\delta\tau$, $\varphi\to \varphi+\delta\varphi$, with
\be\label{fred}
\delta \sigma^i=\p_\mu\sigma^i\int d\sigma^0\, b\,\delta u^\mu,\qquad
 \delta\tau={1 \ov \beta} \delta \beta ,\qquad \delta\varphi=\int d\sigma^0\, b\, \delta\mu
 \ee
which result in %\footnote{Using $X^\mu$ as dynamical variables, the first equation in (\ref{fred}) corresponds to $X^\mu\to X^\mu+\delta X^\mu$, where \be
%\delta X^\mu=\int d\sigma^0\, b\,\delta u^\mu .\ee}
\be\label{fred1}
u^\mu\to u^\mu+\delta u^\mu, \qquad \beta \to \beta +\delta \beta , \qquad \mu\to \mu+\delta\mu \ .
\ee
%$\delta u^\mu,\ \delta \beta$ and $\delta\mu$ should contain at least one derivatives and even number of $a$-fields.
%More explicitly,
$\de u^\mu, \de \beta, \de \mu$ can be expanded in terms of number of derivatives and $a$-fields, e.g.
\be
\de u^\mu = \de u^\mu_0 + i \de u^\mu_1 + \de u^\mu_2 + \cdots
\ee
where $\de u^\mu_n$ contains $n$ factors of $G_{a \mu M}$. $\de u^\mu_0$ starts at first derivatives while $\de u^\mu_k$ with $k >1$ start at zeroth derivatives. Similarly with $\de \beta $ and $\de \mu$.
Under (\ref{fred1}), $\sL_1$ is invariant and
\bega \label{frc}
\sL_r \to \sL_r + \de_r \sL_1 , \quad  \de_r \sL_1 = E_{a\mu}\delta u^\mu+E_{a1} \de \ep_0 +E_{a2}\de n_0
\equiv E_{a \al} \de u^\al , \\
 E_{a \al} \equiv (E_{a \mu}, E_{a1}, E_{a1}) \qquad \de u^\al = (\de u^\mu, \de \ep_0, \de n_0),
\end{gather}
where $\de \ep_0 = \delta \beta \p_\beta \varepsilon_0+\delta\mu\p_\mu\varepsilon_0$, $\de n_0 =\delta \beta\p_\beta n_0+\delta \mu\p_\mu n_0$, and
\be\begin{gathered} E_{a\mu}=\left((\varepsilon_0+p_0) u^\beta G_{a\alpha\beta}+ n_0 C_{a\alpha}\right)\Delta^\alpha_\mu,\\
E_{a1}=\frac 12 u^\mu u^\nu G_{a\mu\nu}+\frac 12 \left(\frac{\p p_0}{\p \varepsilon_0}\right)_{n_0}\Delta^{\mu\nu}G_{a\mu\nu},\quad E_{a2}=u^\mu C_{a\mu}+\frac 12\left(\frac{\p p_0}{\p n_0}\right)_{\varepsilon_0 } \Delta^{\mu\nu}G_{a\mu\nu}\ .\end{gathered}
\label{eaa0}
\ee
We can use~\eqref{frc} to make frame changes.

Below we give some general discussion how we can use~\eqref{iop} and~\eqref{frc} to simplify the action
and the imposing of dynamical KMS condition.

\subsection{Landau frame and generalized Landau frame}\label{sec:lan}

Let us first consider the $O(a)$ Lagrangian $\sL^{(1)} = \ha T^{\mu M} G_{a \mu M}$ which can be written as
\be
\sL^{(1)} =  \frac 12  \le(\vep u^\mu u^\nu +p  \Delta^{\mu\nu}+2 u^{(\mu} q^{\nu)}+\Sig^{\mu\nu} \ri) G_{a\mu\nu}+ \le(n u^\mu+j^\mu\ri) C_{a\mu}+\nabla_\mu H_{a0}^\mu
\ee
where
\be
u_\mu \Sig^{\mu \nu} = u_\mu q^\mu = u_\mu j^\mu = 0, \qquad \De_{\mu \nu} \Sig^{\mu \nu} = 0 \ ,
\ee
and $H_{a0}^\mu$ is an $O(a)$ local expression of fields, i.e. it is linear in $G_{a\mu M}$, and may contain derivatives acting on it, as well as on $u^\mu,\tau$ and $\mu$. Using~\eqref{eaa0}  we can further write $\sL^{(1)}$ as
\be \label{eaa1}
\sL^{(1)}= \mathcal L_{\text{Lan}}^{(1)}+ Q^\al E_{a \al}+\nabla_\mu H_{a0}^\mu %+ E_\al K^\al_a
, \qquad Q^\al = \le(\frac{1}{\varepsilon_0+p_0} q^\mu, \vep, n \ri)
\ee
where
\be \label{finvl0}
\mathcal L_{\text{Lan}}^{(1)}=\ha T_{\rm Lan}^{\mu M} G_{a \mu M} =
\frac 12  \Th_0 \Delta^{\mu\nu}G_{a\mu\nu}
+ \sJ^\mu_0 \De_\mu^\nu C_{a\nu}+\frac 12 \Sig^{\mu \nu} G_{a\mu \nu}\
\ee
and
\be
\Th_0 = p -   \vep \p_\vep p_0 -  n \p_n \ep_0 , \qquad \sJ_0^\mu =  j^\mu -{ n_0  \ov \varepsilon_0+p_0 } q^\mu \ .
\ee
We can further isolate parts of $\Th_0, \sJ^\mu_0, \Sig^{\mu \nu}$ which are proportional to ideal fluid equations of motion
$E_\al$, or to their derivatives. Then~\eqref{eaa1} can be written as
\be \label{linv0}
\sL^{(1)}= \mathcal L_{\text{Lan}}^{(1)}+ Q^\al E_{a \al} + E_\al K^\al_a+\nabla_\mu H_a^\mu
\ee
where $K^\al_a$ is an $O(a)$ local expression of fields, $H_a^\mu$ is an $O(a)$ local expression of $a$- and $r$-type fields, and now $\Th_0, \sJ^\mu_0, \Sig^{\mu \nu}$ in
$\sL_{\text{Lan}}^{(1)}$ only contain tensors which are not related by ideal equations of motion. Terms in $\mathcal L^{(1)}$ that are proportional to derivatives acting on the zeroth order equations have been incorporated in the third term of (\ref{linv0}) upon integrating them by parts. The total derivatives generated in this step have been incorporated in the last term of (\ref{linv0}), together with $H_{a0}^\mu$.\footnote{For example if $\mathcal L_r$ contains a term of the form $A^\nu(\nabla^\mu E_1)G_{a\mu\nu}$ we rewrite it as $-E_1\nabla^\mu(A^\nu G_{a\mu\nu})+\nabla^\mu(E_1 A^\nu  G_{a\mu\nu})$. The first term contributes to $E_\alpha K_a^\alpha$ and the second term contributes to $H_a^\mu$.}

Below the Landau frame Lagrangian refers to this minimal form.
By choosing field redefinitions
\be
\de u^\al = - Q^\al , \qquad \de X^\mu_a = - K^\mu_a
\ee
from~\eqref{iop} and~\eqref{frc} we then obtain $\sL^{(1)} = \mathcal L_{\text{Lan}}^{(1)}+\nabla_\mu H_a^\mu$, where $N_a^\mu$ that appears from doing $a$-field redefinitions (\ref{iop}) has been absorbed in $H_a^\mu$.

We can generalize the above discussion to all orders in the $a$-field expansion. Since $\sL$ contains at least one factor of $G_{a \mu M}$, we can always separate out such a factor and decompose the coefficient of it in terms of tensors parallel and transverse to $u^\mu$, i.e.
\be \label{lanr}
\sL = \frac 12  \le(\mathcal E u^\mu u^\nu +\mathcal P \Delta^{\mu\nu}+2 u^{(\mu}\mathcal Q^{\nu)}+\mathcal S^{\mu\nu} \ri) G_{a\mu\nu}+ \le(\mathcal N u^\mu+\mathcal I^\mu\ri) C_{a\mu}+\nabla_\mu \mathcal H_{a0}^\mu,
\ee
where
\be
u_\mu \sS^{\mu \nu} = u_\mu \sQ^\mu = u_\mu \sI^\mu = 0, \qquad \De_{\mu \nu} \sS^{\mu \nu} = 0 \ ,
\ee
and $\mathcal H_{a0}^\mu$ is a local expression of $a$- and $r$-type fields which is at least $O(a)$. Note $\mathcal E, \mathcal P, \mathcal Q^\mu,\sS^{\mu \nu}$ and $\mathcal N, \mathcal I^\mu$ include terms to all orders in the $a$-field expansion.
%Also in~\eqref{lanr} we can choose to have no derivative acting on $G_{a \mu \nu}$ and $C_{a \mu}$ by integrations by parts.
%unctions of $r$- and $a$-fields, and their derivatives.

Similarly as in the earlier discussion we can further write~\eqref{lanr}  as
\be \label{linv}
\mathcal L = \mathcal L_{\text{Lan}}+ \sQ^\al E_{a \al} + E_\al \sK^\al_a +\nabla_\mu \mathcal H^\mu_a, \qquad
\sQ^\al = \le(\frac{1}{\varepsilon_0+p_0}\mathcal Q^\mu , \mathcal E , \mathcal N \ri),
\ee
where again $\mathcal H_a^\mu$ includes $\mathcal H_{a0}^\mu$ as well as the total derivatives that come from integrating by parts terms that contribute to $E_\alpha \mathcal K^\alpha_a$ above, and $\mathcal L_{\text{Lan}}$ has the form
\be\label{finvl}
\mathcal L_{\text{Lan}}=
\frac 12  \Th \Delta^{\mu\nu}G_{a\mu\nu}
+ \sJ^\mu C_{a\mu}+\frac 12 \mathcal S^{\mu\nu}\Delta_\mu{^\alpha}\Delta_\nu{^\beta} G_{a\al \beta}\
\ee
with
\be
\Th =\mathcal P-   \mathcal E \p_\vep p_0 -  \mathcal N \p_n \ep_0 , \qquad \sJ =  \mathcal I^\mu -{ n_0  \ov \varepsilon_0+p_0 }\mathcal Q^\mu \ .
\ee
In~\eqref{linv} we again have separated possible terms which proportional to ideal fluid equations of motion or to their derivatives.
Note that $\Th_0, \sJ^\mu_0, \Sig^\mu$ correspond respectively to the lowest order terms of  $\Th, \sJ^\mu, \sS^\mu$
in the $a$-field expansion.  By choosing $\de u^\al, \de X^\al_a$ we can then set
\be
\sL = \sL_{\rm Lan} +\nabla_\mu\mathcal H_a^\mu\ ,
\ee
where again we absorbed $N_a^\mu$ in $\mathcal H_a^\mu$. We will refer to~\eqref{finvl} as in the generalized Landau frame.

There is no unique way to write the Lagrangian in the form~\eqref{lanr} since for a cross term of $G_{a \mu \nu}$ and $C_{a \mu}$
one can consider it to be either proportional to  $G_{a \mu \nu}$ or to $C_{a \mu}$.  As a result the generalized Landau frame Lagrangian~\eqref{finvl} is also not unique.
%For example, due to cross terms between $G_{a \mu \nu}$ and $C_{a \mu}$
%the specification of $\sQ^\mu$ and $\sI^\mu$ is not unique; there is freedom to shuffle terms between them.
%Similarly, there is freedom in shuffling terms between $\sP, \sE, \sN$.
Equivalently, from~\eqref{linv} and~\eqref{finvl}, we see that $\sL_r$ is invariant under
\bega \label{amb}
\sJ^\mu \to \sJ^\mu + \Lam_v E_a^\mu , \qquad \Th \to \Th + \Lam_1 E_{a1} + \Lam_2 E_{a2} \\
\sQ^\mu \to \sQ^\mu - (\ep_0 + p_0) \Lam_v C_a^\mu, \quad \sE \to \sE - \ha \Lam_1 \De^{\mu \nu} G_{a \mu \nu} , \quad
\sN \to \sN - \ha \Lam_2 \De^{\mu \nu} G_{a \mu \nu} \ .
\label{amb1}
\end{gather}
where $\Lam_v, \Lam_{1,2}$ are some arbitrary scalar functions (which again has an expansion in $a$-fields).
One could take advantage of the freedom of~\eqref{amb} to make further simplifications.
We will now illustrate this using an explicit example.

\subsection{First order action in the generalized Landau frame}  \label{sec:fir}

 To illustrate the general discussion above more concretely,  let us consider $\sL_2$ of~\eqref{iio}, which we copy here for convenience
  \be \label{lr1}
\mathcal L_2= \ha T^{\mu M}_1 G_{a\mu M}+ {i \ov 4} W^{\mu\nu,MN}_0 G_{a\mu M}G_{a\nu N}
\ee
 $ T^{\mu M}_1$ is given by the first derivative part of~\eqref{1gens}, $W_0^{\mu\nu,MN}$ given by~\eqref{W01}--\eqref{W02}, and their coefficients satisfy~\eqref{1con},~\eqref{11ons}, and~\eqref{1tv}--\eqref{2sc}.
Writing~\eqref{lr1} in the form of~\eqref{lanr} we have (up to the freedom mentioned at the end of last subsection)
\bega
\sE = h_\vep + {i \ov 2} \le(s_{11} u^\lam u^\rho - s_{12} \De^{\lam \rho}  \ri) G_{a \lam \rho} - {i } s_{13} u^\lam C_{a \lam} , \\
\sP = h_p + {i \ov 2} \le(s_{22} \De^{\lam \rho} - s_{12} u^\lam u^\rho  \ri)) G_{a \lam \rho} + {i} s_{23} u^\lam C_{a \lam} \\
\sQ^\mu = q^\mu + i r_{11} u^\lam \De^{\rho \mu} G_{a \lam \rho}  + i r_{12} \De^{\lam \mu} C_{a \lam} \\
\sS^{\mu \nu} = - \eta \sig^{\mu \nu} + 2 i r G_a^{< \mu \nu>} , \quad \sN = h_n +{i \ov 2} (- s_{13} u^\lam u^\rho  + s_{12} \De^{\lam \rho} ) G_{a \lam \rho} + i s_{33} u^\lam C_{a \lam}  \\
\sI^\mu = j^\mu + {i } r_{12} u^\lam \De^{\rho \mu} G_{a \lam \rho}  + i r_{22} \De^{\mu \lam} C_{a \lam}
\end{gather}

In the tensor sector we then find
\be \sS^{\mu\nu}= \eta\left(i\beta^{-1}\Delta^{\alpha\mu}\Delta^{\beta\nu}G_{a\alpha\beta} -\sigma^{\mu\nu}\right)=i\beta^{-1}\eta \Delta^{\alpha\mu}\Delta^{\beta\nu}\tilde G_{a\alpha\beta}(-x),
\ee
where in the above we used the first of (\ref{1tv}) and~\eqref{glkms} (see also~\eqref{u1}).

In the vector sector, using the shift~\eqref{amb} we can choose $\sJ^\mu$ so that its order $O(a)$ term is proportional to $C_{a \mu}$. We then find that $\mathcal J^\mu$ can be written as
\be
 \sJ^\mu =\left(\lambda_{12}+\frac{n_0}{\varepsilon_0+p_0} \lambda_1\right)v_1^\mu-\left(\lambda_2+\frac{n_0}{\varepsilon_0+p_0} \lambda_{12}\right)v_2^\mu+i\beta^{-1}\sigma\Delta^{\mu\alpha}C_{a\alpha},
 \ee
where we used (\ref{1con}), the first of (\ref{11ons}) and (\ref{sio1}). Now using ideal equations of motion~\eqref{0eom} to eliminate $v_{1 \mu}$ we can write the above equation further as
\be
\sJ^\mu=\sigma(i\beta^{-1}\Delta^{\mu\alpha}C_{a\alpha}-v_2^\mu) =i \beta^{-1}\sigma\Delta^{\mu\alpha}\tilde C_{a\alpha}(-x) \ .
\ee

In the scalar sector using the shift in~\eqref{amb} we can choose the $O(a)$ term of $\Th$ to be proportional to $\De^{\mu \nu} G_{a \mu \nu}$, resulting in
\be
\begin{split}
 \Th=& (f_{21}-\p_\varepsilon p_0 f_{11}-\p_n p_0 f_{31})\p \tau-(f_{22}+\p_\varepsilon p_0 f_{12}+\p_n p_0 f_{32})\theta\\
&+(f_{23} - \p_\varepsilon p_0 f_{12}+\p_n p_0 f_{33})\beta^{-1}\p\hat \mu+\frac i2 \beta^{-1} \zeta \Delta^{\mu\nu} G_{a\mu\nu},\end{split}
\ee
where we used (\ref{ze1}). Further using equations of motion we then find
\be
\Th =\zeta\left(\frac i2 \beta^{-1}\Delta^{\mu\nu}G_{a\mu\nu}-\theta\right)=\frac i2 \beta^{-1}\zeta\Delta^{\mu\nu}\tilde G_{a\mu\nu}(-x),
\ee
where to write the $O(a^0)$ part of $\Theta$ we used \eqref{11ons} and~\eqref{1sc}--\eqref{2sc}.

Finally collecting the above results together we find a remarkably simple expression
\be
\sL_2 = \mathcal L_{\text{Lan}}=\frac i4\beta^{-1}\zeta\Delta^{\mu\nu}\tilde G_{a\mu\nu}\Delta^{\alpha\beta} G_{a\alpha\beta}+i\beta^{-1}\sigma \Delta^{\mu\alpha} \tilde C_{a\alpha} C_{a\mu}+\frac i2 \beta^{-1}\eta \Delta^{\mu\alpha}\delta^{\nu\beta}\tilde G_{a\mu\nu} G_{a\alpha\beta},
\ee
where the tilded variables should be evaluated at $-x$. Note that $\mathcal L_{\text{Lan}}$ is manifestly dynamical KMS invariant.

\subsection{Dynamical KMS condition in Landau frame}\label{sec:lffr}

%In this subsection we discuss how field redefinitions modify the dynamical KMS condition.
%to impose the dynamical KMS condition in the generalized Landau frame.
%In particular, to simplify the analysis it would be ideal if we are able to impose the dynamical KMS condition in the Landau frame.

%To see how a Lagrangian in the generalized Landau frame transforms under dynamical KMS transformation let us trace how field redefinitions in going to the generalized Landau frame

%where $\tilde \sL^{(0)}$ denotes terms in $\tilde \sL$ which have no $a$-field dependence. Note that
%\be
%\tilde \sL^{(0)} = \tilde \sL_0^{(0)} + \tilde \sL_r^{(0)}
%\ee
%and the corresponding condition
%\be
%\tilde \sL_0^{(0)}  = \p_\mu V_{(0,0)}^\mu
%\ee

Starting with a dynamical KMS invariant action, after field redefinition changes~\eqref{iop} and~\eqref{frc}, the resulting action is in general {\it no longer} dynamical KMS invariant. This is fine
as the  generating functional $W [g_1,A_1;g_2,A_2]$ of~\eqref{qft0} should be invariant under such field redefinitions and remains KMS invariant.\footnote{In evaluating the path integral beyond tree-level, one will have to be careful about potential changes in the integration measures due to such field redefinitions.} Thus it comes as a pleasant surprise that the action involving $T_1^{\mu M}$ can be written in a manifest dynamical KMS invariant form in the generalized Landau frame. It would be interesting to explore
whether this happens at all odd derivative orders.

In this subsection to prepare for the explicit second order analysis in Sec.~\ref{sec:conf} we
give some general discussion on how  the dynamical KMS condition imposes constraints on $T^{\mu M}_{\rm Lan}$   at even derivative orders.

We start with the Lagrangian obtained from~\eqref{w4} for which the dynamical KMS condition is satisfied at $O(a^k)$ for $k\geq 1$ in the $a$-field expansion. One only needs to impose~\eqref{gg0}, which we copy here for convenience
\be \label{gg1}
\le(\tilde \sL_{2n+1}\ri)_0 = \le(\widetilde{\sL^{(1,2n)}} + \widetilde{\sL^{(2,2n-1)}} + \cdots
+ \widetilde{\sL^{(2n+1,0)}} \ri)_0 = i \nab_\mu V^\mu_{(0,2n)} \ , \quad n=1, 2, \cdots  \ .
\ee
Field redefinitions for different $n$'s can be treated independently, so let us now consider a specific $n$.
Consider a field redefinition which takes  $\sL_r^{(1,2n)}$ to the Landau frame, i.e
\be
\sL_{\rm Lan}^{(1,2n)} = \sL^{(1,2n)} - E_{a \al} \de u^\al -\de X^\al_a E_\al +\nabla_\mu H_a^\mu \ .
\ee
 $\de u^\al$ is of $O(a^0)$ and contains $2n$ derivatives, while $\de X^\al_a$ is $O(a)$ with $2n-1$ derivatives.
Note none of the other $\sL^{(m,2n+1-m)}$ in~\eqref{gg1} is affected by such redefinitions.

Under a dynamical KMS transformation
\be \label{hrn}
\widetilde{\sL_{\rm Lan}^{(1,2n)}}  = \widetilde{\sL^{(1,2n)}} - (E_{a \al} +  i \beta E_\al) \de u^\al - E_\al (\de X^\al_a + i \beta \de Y^\al)+\nabla_\mu(H_a^\mu+i Z^\mu)
\ee
where we have used that $\widetilde{\de u^\al} (-x) = \de u^\al (x), \tilde E_\al (-x) = - E_\al (x)$, and
\be
\tilde E_{a \al} (-x) = E_{a \al} (x) + i \beta E_\al (x)  \ .
\ee
Since $\de X^\al_a$ and $H_a^\mu$ are $O(a)$, under a dynamical KMS transformation they must have the form
\be
\widetilde{\de X^\al_a} (-x) = - (\de X^\al_a (x) + i \beta \de Y^\al (x)),\quad
\tilde H^\mu_a(-x)=-(H_a^\mu(x)+i Z^\mu(x))
\ee
for some  $\de Y^\al$ and $Z^\mu$ which are $O(a^0)$, and the overall minus sign on the right hand side is due to that $\de X^\al_a$ and $H_a^\mu$ have odd number of derivatives.
Note that the tilde operation in~\eqref{hrn} should be understood in the sense as described below~\eqref{inVV}.
One can readily check that~\eqref{hrn} does not affect the dynamical KMS invariance at $O(a)$,  which also follows from that dynamical KMS transformation at order $O(a)$ does not constrain even derivative terms in $\sL^{(1)}$.

Now plugging~\eqref{hrn} into~\eqref{gg1} we find that
\be \label{lan1}
\le(\widetilde{\sL_{\rm Lan}^{(1,2n)}} \ri)_0 + \sum_{k=2}^{2n+1} \le( \widetilde{\sL_r^{(k, 2n+1-k)}} \ri)_0 =i  \nab_\mu \bar V^\mu_{2n} +   i \beta E_\al (\de u^\al + \de Y^\al)\ ,
\ee
where $\bar V^\mu_{2n}= V^\mu_{(0,2n)}+Z^\mu$. Note that from~\eqref{finvl0}
\be \label{lan2}
\le(\widetilde{\sL_{\rm Lan}^{(1)}} \ri)_0  =  i \beta \le( \Th_0 \th  +\sJ^\mu_0 v_{2 \mu} + \ha  \Sig^{\mu \nu}_0 \sig_{\mu \nu} \ri)   \ .
\ee

In Appendix \ref{app:dec} we prove that at any derivative order $m$ and tensor rank $k$, one can always choose a set of basis of the form $\{v_{1}^{\mu_1\cdots \mu_k},\dots,v_{p}^{\mu_1\cdots \mu_k},w_{1}^{\mu_1\cdots \mu_k},\dots, w_{q}^{\mu_1\cdots \mu_k}\}$, where $v_{i}^{\mu_1\cdots \mu_k}$ are not related by the ideal fluid equations (\ref{0eom}), and $w_{s}^{\mu_1\cdots \mu_k}$ contain at least one factor of (\ref{0eom}) or their derivatives. For later reference, we call the first ones $v$-type tensors, and the second ones $w$-type tensors.

Now with the definition below~\eqref{linv0} for the Landau frame, we can then write~\eqref{lan2} {\it solely} in terms of $v$-type tensors, i.e.
 \be
\le(\widetilde{\sL_{\rm Lan}^{(1)}} \ri)_0 = \sum_i a_i v_i
\ee
where $a_i$ are functions of $\tau$ and $\hmu$. Similar we can write
\bega
\sum_{k=2}^{2n+1} \le( \widetilde{\sL_r^{(k, 2n+1-k)}} \ri)_0 = \sum_i b_i v_i + \sum_s c_i w_s  \\
 \nab_\mu\bar V^\mu_{2n} =  \le(\nab_\mu \bar V^\mu_{2n}\ri)_{\rm min}  + \sum_i d_s w_s
\end{gather}
where $ \le(\nab_\mu \bar V^\mu_{2n}\ri)_{\rm min}$ contains only $v_i$ terms. Equation~\eqref{lan1} can thus be written as
\be \label{lan4}
\le(\tilde \sL_{\rm Lan}^{(1,2n)} \ri)_0 + \sum_i b_i v_i   =  \le(\nab_\mu \bar V^\mu_{2n}\ri)_{\rm min}
\ee
and
\be
 \sum_s c_s w_s =  \sum_s d_s w_s +   i \beta E_\al (\de u^\al + \de Y^\al) \ .
 \ee
 Equation~\eqref{lan4} implies when imposing the dynamical KMS condition on $\sL_{\rm Lan}^{(1,2n)}$ we can
 set all terms proportional to ideal equations of motion to zero. As we will see in next section, this provides a great simplification.

\subsection{Entropy current in Landau frame}

In this subsection we show that the entropy current in Landau frame satisfies the local second law up to terms that vanish on the ideal fluid equations of motion. To this aim, we need to carefully track the steps in going from (\ref{lan1}) to the equation of the entropy divergence. Recall that eq. (\ref{lan1}) is a manipulation of the $(2n+1)$th derivative order part of (\ref{nnm}). Summing (\ref{lan1}) to the lower derivative orders of (\ref{nnm}) we then find
\be\label{lan1a}\sum_{j=0}^{2n}\le(\widetilde{\sL_{\rm Lan}^{(1,j)}} \ri)_0 + \sum_{j=0}^{2n}\sum_{k=2}^{j+1} \le( \widetilde{\sL_r^{(k, j+1-k)}} \ri)_0 =i  \nab_\mu  \mathcal V^\mu_{2n} +   i \beta E_\al (\de u^\al + \de Y^\al)\ ,\ee
where $\mathcal V^\mu_{2n}=\sum_{j=0}^n \bar V^\mu_{2j}$, and where we assume that the Lagrangian at derivative order lower than $2n$ is already in Landau frame. Using (\ref{finvl0}) we express the first term in (\ref{lan1a}) in terms of the stress tensor and the charge current,
\be\label{lan1b}\frac 12 T^{\mu M}_{\text{Lan}} \Phi_{r\mu M}-i \sum_{j=0}^{2n}\sum_{k=2}^{j+1} \le( \widetilde{\sL_r^{(k, j+1-k)}} \ri)_0 =  \nab_\mu \mathcal V^\mu_{2n} +    \beta E_\al (\de u^\al + \de Y^\al)\ ,\ee
integrating by parts,
\be \label{kmsa0}\nabla_\mu \left(\mathcal V_{2n}^\mu-T^{\mu M}_{\text{Lan}}\beta_M\right)=-\nabla_\mu T^{\mu M}_{\text{Lan}}\beta_M-i\sum_{j=0}^{2n}\sum_{k=2}^{j+1} \le( \widetilde{\sL_r^{(k, j+1-k)}} \ri)_0- \beta E_\al (\de u^\al + \de Y^\al)\ ,\ee
where
\be \nabla_\mu T^{\mu M}_{\text{Lan}}\beta_M=(\nabla_\mu T_{\text{Lan}}^{\mu\nu}+J_{\text{Lan}}^\mu F_\mu^{\ \nu})\beta_\nu+\nabla_\mu J_{\text{Lan}}^\mu \hat \mu,\quad \beta_M=(\beta_\mu,\hat\mu)\ .\ee
The $O(a^0)$ part of the exact equations of motion for the Landau frame Lagrangian is
\be \label{eome}\nabla_\mu T_{\text{Lan}}^{\mu\nu}=-J_{\text{Lan}}^\mu F_\mu^{\ \nu},\quad \nabla_\mu J^\mu_{\text{Lan}}=0,\ee
and imposing the above, eq. (\ref{kmsa0}) becomes
\be \label{kmsa1}\nabla_\mu \left(\mathcal V_{2n}^\mu-T^{\mu M}_{\text{Lan}}\beta_M\right)=-i\sum_{j=0}^{2n}\sum_{k=2}^{j+1} \le( \widetilde{\sL_r^{(k, j+1-k)}} \ri)_0- \beta E_\al (\de u^\al + \de Y^\al)\ .\ee
Eq. (\ref{kmsa1}) has the same form as eq. (3.13) in \cite{GL}, except that in (\ref{kmsa1}) we have an additional term on the RHS which is proportional to the ideal fluid equations of motion, whereas the first term was shown in \cite{GL} to be always non-negative. This shows that the entropy current $S_{\text{Lan}}^\mu$ in Landau frame,
\be S_{\text{Lan}}^\mu\equiv \mathcal V^\mu- T_{\text{Lan}}^{\mu M}\beta_M=(p_0+\varepsilon_0)\beta^\mu- J_{\text{Lan}}^\mu\hat\mu+\sum_{j=1}^n \bar V^\mu_{2j},\ee
is guaranteed to satisfy the local second law at all derivative orders, up to terms that vanish on the ideal equations of motion.

\section{Conformal fluids at second order in derivatives} \label{sec:conf}

In this section we consider the action for a conformal fluid which has some new elements.
We will also work out explicitly the corresponding entropy current to second order in derivative expansion using~\eqref{s0} and show that it reproduces previous results.

\subsection{Conformal fluids} \label{sec:7a}

 For a conformal system the generating functional~\eqref{pager1} should in addition be invariant under independent Weyl scalings of
 two metrics
\be
\label{conW} W[g_{1\mu\nu},A_{1\mu},g_{2\mu\nu},A_{2\mu}] =W[e^{2\lambda_1}g_{1\mu\nu},A_{1\mu},e^{2\lambda_2}g_{2\mu\nu},A_{2\mu}]
\ee
where $\lambda_s(x)$ are scalars, and $s=1,2$.
For this purpose it is convenient to introduce Weyl invariant fluid spacetime metrics
\be \label{hnew}
\hat h_{sab} (\sig)  ={1 \ov \beta_s^2}  h_{sab}(\sigma) , \quad \beta_{s} = \beta_0 e^{ \tau_{s} (\sig)}, \quad
\tau_1=\tau+\frac 12 \tau_a, \quad \tau_2=\tau-\frac 12 \tau_a
\ee
where $h_{sab}$ was defined in (\ref{hdef1}) and $\tau_a$ is defined form the determinants of $h_{1ab}$ and $h_{2ab}$,
\be
e^{d\tau_a}=\sqrt{\det(h_1 h_2^{-1})} \ .
\ee
Under Weyl scalings $g_{s \mu \nu} (X) \to e^{2 \lam_s (X)} g_{s \mu \nu}$,  $\tau_a$ transforms as $\tau_a\to \tau_a+ \lambda_1 (\sig)- \lambda_2 (\sig)$, and $\hat h_{s ab}$ are invariant if $\tau$  transforms as
\be \label{weyl}
2 \tau(\sig) \to 2 \tau (\sig) + \lam_1 (\sig)+ \lam_2 (\sig), \qquad \lam_s (\sig) = \lam_s (X (\sig)) \ .
\ee
Equation~\eqref{conW} can be satisfied if we require the action to depend only on $\hat h_{sab}$ and $B_{sa}$, i.e.
\be \label{cftf}
{\rm conformal \; fluids}: \quad I = I [\hat h_1, B_1; \hat h_2; B_2] \ .
 \ee
All the other conditions discussed in the Introduction section remain the same.

One can immediately write down the action either in fluid spacetime or physical spacetime
in parallel with earlier discussions using $\hat h_{s ab}$ in place of $h_{s ab}$
and dropping any explicit $\tau$-dependence.
For illustration we will consider a {\it neutral} fluid. In Appendix~\ref{app:conf1} we discussion the formulation of the action
in the fluid spacetime with a finite $\hbar$ (before imposing dynamical KMS symmetry).
Here we concentrate on the classical limit.

In the $\hbar \to 0$ limit using (\ref{hnew}), one finds
\be
\hat h_{1ab}= \hat h_{ab}(\sigma)+\frac{\hbar}2\hat h_{ab}^{(a)}
,\ee
with
\begin{gather}
\hat h_{ab}(\sigma)\equiv \p_a X^\mu \p_b X^\nu \hat g_{\mu\nu}(X),\qquad
\hat h_{ab}^{(a)}=\p_a X^\mu \p_b X^\nu \hat G_{a\mu\nu}(X),\\
\hat g_{\mu\nu}\equiv \beta^{-2}(x)  g_{\mu\nu},\qquad
\hat G_{a\mu\nu}\equiv \left(\beta^{-2}(x)  g_{a\mu\nu}+\mathcal L_{X_a}\hat g_{\mu\nu}\right)_{\text{traceless}}
= \beta^{-2}(x)\left(G_{a\mu\nu}\right)_{\text{traceless}},
\end{gather}
where $\beta(x)$ is defined as in~\eqref{tay0}, and the traceless part of a tensor $A$ is defined as
\be
\left(A_{\mu\nu}\right)_{\text{traceless}}=A_{\mu\nu}-\frac 1d A_{\alpha\beta}\hat g^{\alpha\beta}\hat g_{\mu\nu}
\ee
with $\hat g^{\mu\nu}$  the inverse of $\hat g_{\mu\nu}$. The action in physical space-time will now depend on
$ \beta^\mu, \hat g_{\mu\nu},  \hat G_{a\mu\nu}$ with
\be
\beta^\mu=\frac 1{\hat b}\p_0 X^\mu,\qquad \hat b=\sqrt{-\hat g_{\mu\nu}\p_0 X^\mu\p_0 X^\nu} \ . %,\quad \hat \mu=\beta^\mu B_\mu.
\ee
For spacetime derivatives we will always use the covariant derivative $\hat \nabla_\mu$ associated with $\hat g_{\mu\nu}$. The dynamical KMS transformations~\eqref{glkms} now have the form
\be
\begin{gathered}\label{hatg}
\beta^\mu(x)\to \beta^\mu(-x),\quad
\hat g_{\mu\nu}(x) \to \hat g_{\mu\nu}(-x),\quad
\hat G_{a\mu\nu}(-x)\to \hat G_{a\mu\nu}(x)+2\left(\hat\nabla_{(\mu}\beta_{\nu)}\right)_{\text{traceless}}(x) \ ,
\end{gathered}
\ee
where we use $\hat g_{\mu\nu}$ and its inverse to raise and lower indices. In particular, note that in terms of usual velocity field $u^\mu$,
\be
\beta^\mu = \beta u^\mu , \qquad  \beta_\mu = \hat g_{\mu\nu}\beta^\nu = \beta^{-1}(x) g_{\mu\nu} u^\nu, \qquad
\beta^\mu \beta_\mu = \hat g_{\mu \nu} \beta^\mu \beta^\nu = -1 \ .
\ee
%\footnote{In particular, note that $\beta_\mu = \hat g_{\mu\nu}\beta^\nu = \beta^{-1}(x) g_{\mu\nu} u^\nu$.}

Now the action can be written as
\be
I_{\rm hydro} = \int d^d x \, \sqrt{-\hat g} \, \hat \sL , \qquad \hat \sL = \beta^d \sL
\ee
with
%It is convenient to write the Lagrangian (\ref{acttja}) as
\be\label{conflag}
\hat \sL= \ha  \hat T^{\mu \nu}\hat G_{a\mu \nu}+ {i \ov 4} \hat W^{\mu\rho,\nu\sigma} \hat G_{a\mu \nu}\hat G_{a\rho\sigma}+ {1 \ov 8} \hat Y^{\mu\rho\alpha,\nu\sigma\beta}\hat G_{a\mu \nu}\hat G_{a\rho\sigma}\hat G_{a\alpha\beta}+\cdots \ .
\ee
where $\hat T^{\mu\nu}$, $\hat W^{\mu\nu, \lam \rho}$ and $\hat Y^{\mu\nu\rho, \lam\sig \de }$ are functions of $\beta^\mu, \hat g_{\mu\nu}$,  their derivatives, and derivative operators acting on $\hat G_{a\mu M}$. Note that since $\hat G_{a\mu\nu}$ is traceless, the trace components $\hat T^{\mu\nu}\hat g_{\mu\nu},\ \hat X^{\mu\rho, \nu\sigma}\hat g_{\mu\nu},\ \dots$ decouple from the action, and we shall thus take such components to be zero in what follows. From (\ref{hatg}), these hatted tensors are related to the un-hatted ones through $T^{\mu\nu}=\beta^{-(d+2)}(x)\hat T^{\mu\nu}$, $W^{\mu\rho,\nu\sigma}=\beta^{-(d+4)}(x)\hat W^{\mu\rho,\nu\sigma} $ and $Y^{\mu\rho\alpha,\nu\sigma\beta}=\beta^{-(d+6)}(x)\hat Y^{\mu\rho\alpha,\nu\sigma\beta}$.

To second order in derivative expansion, the dynamical KMS conditions can be imposed using~\eqref{uio0}--\eqref{004}.
At zeroth and first order the analysis are the same as before and we find
\begin{align} \hat T_0^{\mu\nu}=&\hat p_0((d-1)\beta^\mu \beta^\nu+\hat \Delta^{\mu\nu})\\
\label{hT1}\hat T_1^{\mu\nu}=&-(d-1)\hat f_{22}\hat\theta \beta^\mu\beta^\nu-\hat f_{22}\hat\theta\hat\Delta^{\mu\nu}-2\hat\lambda_1\beta^{(\mu}\hat\p \beta^{\nu)}-\hat\eta \hat \sigma^{\mu\nu}\\
\hat W_0^{\mu\alpha,\nu\beta}=&(d-1)^2 \hat s_{22}\beta^\mu \beta^\nu\beta^\alpha\beta^\beta+\hat s_{22}\hat\Delta^{\mu\nu}\hat\Delta^{\alpha\beta}+(d-1) \hat s_{22}(\beta^\mu\beta^\nu\hat \Delta^{\alpha\beta}+\beta^\alpha\beta^\beta\hat \Delta^{\mu\nu})\\
&+2\hat r_{11}(\beta^\mu\beta^{(\alpha}\hat\Delta^{\beta)\nu} +\beta^\nu\beta^{(\alpha}\hat\Delta^{\beta)\mu})+4\hat r\hat \Delta^{\alpha<\mu}\hat\Delta ^{\nu>\beta},\label{hW02}
\end{align}
where all coefficients  are constants,
\be
\hat\Delta^{\mu\nu}=\hat g^{\mu\nu}+\beta^\mu\beta^\nu,\qquad
\hat \sigma_{\mu\nu}=2\hat\nabla_{<\mu}\beta_{\nu>},\qquad \hat \theta= \hat\nabla_\mu\beta^\mu,\qquad  \hat\p=\beta^\mu\hat\nabla_\mu,
\ee
and $A^{<\mu\nu>}$ denotes the symmetric transverse traceless part of a tensor $A^{\mu \nu}$, i.e.
\be
A^{<\mu\nu>} \equiv
A^{(\alpha\beta)}\hat\Delta_\alpha^{\ \mu}\hat\Delta_\beta^{\ \nu}-\frac 1{d-1}A^{\alpha\beta}\hat\Delta_{\alpha\beta}\hat\Delta^{\mu\nu} \ .
\ee
Invariance of $\hat {\mathcal L}$ under (\ref{hatg}) gives the following relations among coefficients in (\ref{hT1})-(\ref{hW02})
\be
\hat r=\frac 12 \hat\eta,\qquad \hat r_{11}=\hat \lambda_1,\qquad \hat s_{22}=\hat f_{22},
\ee
which is the conformal limit of (\ref{1tv})--(\ref{2sc}). Interestingly, equations (\ref{1con}) and (\ref{11ons}) are automatically satisfied in the conformal case.

\subsection{Analysis at second order}

To obtain the explicit form of the entropy current~\eqref{s0} to second derivative order,
in this subsection we work out the constraint~\eqref{004} for a conformal fluid for which it becomes
\be \label{004a}
\hat T^{\mu \nu }_2 \hat\nabla_\mu\beta_\nu + {1 \ov 2} \hat Y^{\mu \rho\alpha,\nu\sigma\beta}_0  \hat\nabla_\mu\beta_\nu  \hat\nabla_\rho\beta_\sigma  \hat\nabla_\alpha\beta_\beta
= \hat\nabla_\mu\hat V^\mu_{(0,2)}  \ ,
\ee
where $\hat V_{(0,2)}^\mu=\beta^d(x) V_{(0,2)}^\mu$, and we have used $\nabla_\mu V_{(0,2)}^\mu = \beta^{-d}(x)\hat\nabla_\mu \hat V_{(0,2)}^\mu$.
To simplify the analysis we will apply the discussion of Sec. \ref{sec:lffr}: we will go to Landau frame and set to zero terms that vanish on the ideal fluid equations of motion. Note in the conformal case the ideal fluid equations $\nabla_\mu T_0^{\mu\nu}=0$ can be written as
\be \label{ceom}
\hat\p\beta_\mu=0, \qquad \hat\theta=0 \ .
\ee

The explicit expression of $\hat T_2^{\mu\nu}$ in Landau frame is
\be\label{Tmn1}
\hat T^{\mu\nu}_{2}=f_1 \hat R^{<\mu\nu>}+f_2 \hat \sigma^{<\mu}_{\ \alpha}\hat\sigma^{\nu>\alpha}+f_3 \hat\omega^{<\mu}_{\ \ \ \alpha}\hat\omega^{\nu>\alpha}+f_4 \hat\sigma^{<\mu}_{\ \alpha}\hat\omega^{\nu>\alpha}+f_5(\hat\p \hat\sigma)^{<\mu\nu>},
\ee
with
\be
\hat\omega^{\mu\nu} \equiv -2\hat\Delta^{\mu\alpha}\hat\Delta^{\nu\beta}\hat \nabla_{[\alpha}\beta_{\beta]}, \quad
(\hat\p \hat\sigma)^{<\mu\nu>} \equiv \hat \Delta^{\mu\alpha}\hat\Delta^{\nu\beta}\hat\p (\hat \sigma_{\alpha\beta})-\frac 1{d-1}\hat \Delta^{\mu\nu}\hat \Delta^{\alpha\beta}\hat \p(\hat \sigma_{\alpha\beta}) %= \hat \Delta^{\mu\alpha}\hat\Delta^{\nu\beta}\hat\p (\hat \sigma_{\alpha\beta}),
\ee
where $\hat R_{\alpha\beta}$ is the Ricci tensor of $\hat g_{\mu\nu}$, and
$f_1,\ f_2,\ \dots$ are constant, and where again we neglected terms that vanish on (\ref{ceom}), such as  $\hat\theta\hat\sigma^{\mu\nu}$. Similarly after writing down the most general tensor form of $ \hat Y^{\mu \rho\alpha,\nu\sigma\beta}_0 $, contracting it with $\hat \nab_\mu \beta_\nu$, and setting to zero all terms proportional to~\eqref{ceom}, we find for the second term of~\eqref{004a} only one term survives
\be \label{Y01}
{1 \ov 16} \hat Y^{\mu \rho\alpha,\nu\sigma\beta}_0 \hat\sigma_{\mu\nu}\hat\sigma_{\rho\sigma}\hat\sigma_{\alpha\beta}=\frac 1{16} h_1\hat \sigma^\mu_{\alpha}\hat\sigma^{\nu\alpha}\hat\sigma_{\mu\nu} \
\ee
with $h_1$ a constant. Finally, the most general expression for $\hat V^\mu_{(0,2)}$ is\footnote{Note that below we are identifying $V_{(0,2)}^\mu$ with $\bar V_{2n}^\mu$ introduced below (\ref{lan1}).}
\be
\hat V_{(0,2)}^\mu = (c_1\hat R+c_2\hat\sigma^2+c_3\hat\omega^2) \beta^\mu + v_1 \hat\nabla_\nu\hat\omega^{\mu\nu}+v_2\left(\hat R^{\mu\nu}-\frac 12 \hat g^{\mu\nu}\hat R\right)\beta_\nu,\label{V02}
\ee
where $\omega^2=\omega_{\mu\nu}\omega^{\mu\nu}$ and  $c_1,\ c_2,\ \dots$ are constant.
In writing down  (\ref{Tmn1})-(\ref{V02}) we have used the identities of Appendix \ref{app:id} which
guarantee that these are the most general second order expressions.

Plugging (\ref{Tmn1}),(\ref{Y01}) and (\ref{V02}) into~\eqref{004a}, and using again the identities of Appendix \ref{app:id}, we find
\be\label{div0}\begin{gathered}\frac 14 f_5\hat\p(\hat\sigma^2)+\frac 12 f_3\hat \omega^\mu_{\ \alpha}\hat\omega^{\nu\alpha}\hat\sigma_{\mu\nu}+\left(\frac 12 f_2+\frac 1{16} h_1\right)\hat \sigma^\mu_{\alpha}\hat\sigma^{\nu\alpha}\hat\sigma_{\mu\nu}+\frac 12 f_1\hat R^{\mu\nu}\hat\sigma_{\mu\nu}\\
=c_1\hat \p \hat R+c_2\hat\p(\hat\sigma^2)-2c_3\hat \omega^\mu_{\ \alpha}\hat\omega^{\nu\alpha}\hat\sigma_{\mu\nu}+\frac 12 v_2\hat R^{\mu\nu}\hat\sigma_{\mu\nu},\end{gathered}\ee
where in evaluating the divergence $\hat\nabla_\mu \hat V^\mu_{(0,2)}$ we again neglected terms that vanish on the ideal fluid equations. Eq. (\ref{div0}) gives the relations
\be\label{rels}
c_1=0,\qquad c_2=\frac 14 f_5,\qquad c_3=-\frac 14 f_3,\qquad v_2=f_1,\qquad f_2=-\frac 18 h_1 \ ,
 \ee
Note that $v_1$ does not appear in the above relations as in (\ref{V02}) it multiplies a term of zero divergence.\footnote{This comes from the identity $[\hat \nabla_\mu,\hat\nabla_\nu]\hat\omega^{\mu\nu}=2\hat R_{\mu\nu}\hat\omega^{\mu\nu}=0$, which holds without imposing the ideal fluid equations.} Equation~(\ref{rels})
gives
\be
\label{v02a} V_{(0,2)}^\mu=\beta^{-d}\hat V_{(0,2)}^\mu=\frac 14 (f_5\hat\sigma^2-f_3\hat\omega^2) \beta^\mu + v_1 \hat\nabla_\nu\hat\omega^{\mu\nu}+f_1\left(\hat R^{\mu\nu}-\frac 12 \hat g^{\mu\nu}\hat R\right)\beta_\nu  \ .
\ee
%together with the constraint
%\be\label{g2f2} f_2=-\frac 18 h_9.\ee

In the above analysis we did not find the relation
\be \label{1ya}
f_5+f_4-2 f_2=0
\ee
which was observed as a universal relation in holographic theories dual to Einstein gravity~\cite{Haack:2008xx}.
 Equation~\eqref{1ya} was moreover found to be present  in the first order correction in various higher derivative
theories~\cite{Shaverin:2012kv,Grozdanov:2014kva}, but  fail non-perturbatively
in Gauss-Bonnet coupling~\cite{Grozdanov:2014kva,Grozdanov:2015asa} (which was independently verified at second order in Gauss-Bonnet coupling in~\cite{Shaverin:2015vda}). Our conclusion is consistent with the discussion in~\cite{Grozdanov:2014kva} that such a relation cannot hold universally  in hydrodynamics.

\subsection{Entropy current at second order}

From~\eqref{s0} the expression for the entropy current at second order, for a conformal neutral fluid, is
\be \label{90}
S_2^\mu=V_{(0,2)}^\mu-T_2^{\mu\nu}\beta_\nu = V_{(0,2)}^\mu
%=\frac 14 (f_5\hat\sigma^2-f_3\hat\omega^2) \beta^\mu + v_1 \hat\nabla_\nu\hat\omega^{\mu\nu}+f_1\left(\hat R^{\mu\nu}-\frac 12 \hat g^{\mu\nu}\hat R\right)\beta_\nu,
\ee
where the second equality follows from that $T_2^{\mu\nu}$ is in Landau frame.
The expression~\eqref{90} with $V_{(0,2)}^\mu$ given by~\eqref{v02a} agrees precisely with the expression previously
given in~\cite{bhat,Romatschke:2009kr}, except that with the method of \cite{bhat,Romatschke:2009kr} $c_2$ was undetermined (see also \cite{Bhattacharyya:2014bha}).
%where we used (\ref{v02a}) and we assumed $T_2^{\mu\nu}u_\nu=0$.
%Taking the divergence of~\eqref{90} we find that (\HL{equation below appears to be contradictory with the second equality of~\eqref{90}})
Taking the divergence of the total entropy current (\ref{s0}), using the equations of motion (\ref{eom}), (\ref{uio0}), (\ref{003}) and (\ref{004}), the third order part of the entropy divergence is
\be (\nabla_\mu S^\mu)_3 =-\frac 12 T_2^{\mu\nu} \Phi_{r\mu\nu}+\nabla_\mu V_{(0,2)}^\mu
= \frac 1{16}Y_0^{\mu\rho\alpha,\nu\sigma\beta}\Phi_{r\mu\nu} \Phi_{r\rho\sigma}\Phi_{r\alpha\beta}  %\ .
\ee
%Using (\ref{004}) and the first in (\ref{uio0}),
%\be \nabla_\mu S^\mu=\frac 14 W_0^{\mu\rho,\nu\sigma}\Phi_{r\mu\nu}\Phi_{r\rho\sigma} +\frac 1{16}Y_0^{\mu\rho\alpha,\nu\sigma\beta}\Phi_{r\mu\nu} \Phi_{r\rho\sigma}\Phi_{r\alpha\beta}.\ee
%Using again the ideal equations of motion we can write explicitly
which leads to
\be
(\nabla_\mu S^\mu)_3 = % \frac 1{2 T}\eta \sigma^{\mu\nu}\sigma_{\mu\nu}
-\ha T^{d-3} f_2 \sigma^\mu_{\alpha}\sigma^{\nu\alpha}\sigma_{\mu\nu} \ .
\ee
Note that the right hand side of the above equation does not have a definite sign. This term is subleading in derivative expansion compared with the first term in~\eqref{entp}. Altogether, up to third order we have
\be
\nabla_\mu S^\mu =  \ha T^{-1}\eta \sigma^{\mu\nu}\sigma_{\mu\nu} -\ha T^{d-3} f_2 \sigma^\mu_{\alpha}\sigma^{\nu\alpha}\sigma_{\mu\nu} \label{divs4}
\ee
As shown in \cite{GL}, the entropy divergence can always be written as a square, up to higher derivative terms. Applying the algorithm constructed in \cite{GL} we write (\ref{divs4}) as
\be
\nabla_\mu S^\mu  =  \ha T^{-1}\eta\left( \sigma^{\mu\nu}\sigma_{\mu\nu} -\ha T^{d-2} \frac{f_2}{\eta} \sigma^\mu_{\alpha}\sigma^{\nu\alpha}\right)^2-\frac 18 T^{2d-5}\frac{f_2^2}{\eta}\sigma^\mu_{\alpha}\sigma^{\alpha}_\beta \sigma^\beta_\gamma\sigma^\gamma_\mu
\ee
the last term is fourth order in derivatives, hence it can be neglected, and we are left with a non-negative divergence.

\section{Conclusions and Discussions} \label{sec:dis}

In this paper we further developed the fluctuating hydrodynamics proposed in~\cite{CGL} in a number of directions.
We first elucidated the structure of the hydrodynamic action in the classical limit, which enables a transparent formulation of the action in physical spacetime in the presence of arbitrary external fields. It also makes connections to some of the earlier work~\cite{Dubovsky:2005xd,Dubovsky:2011sj,Endlich:2012vt,Dubovsky:2011sk,Endlich:2010hf,
Nicolis:2011ey,Nicolis:2011cs,Delacretaz:2014jka,Geracie:2014iva,Grozdanov:2013dba,
Kovtun:2014hpa,Harder:2015nxa} clearer.
We then proposed a dynamical KMS symmetry which ensures local equilibrium. The dynamical KMS symmetry is physically equivalent  to the previously proposed local KMS condition in the classical limit,  but is more convenient to implement and more general. It should be applicable to any states in local equilibrium rather than just thermal density matrix perturbed by external sources.  We then discussed making frame choices using field redefinitions, which can be used to significantly simplify the action and the imposition of the dynamical KMS symmetry. We discussed how to go to the Landau frame and generalized Landau frame.
 Finally we proposed a formulation for a conformal fluid, which requires introducing some new elements.
 We then worked out the explicit form of the entropy current to second order in derivatives for a neutral conformal fluid using the
 method of~\cite{GL}. The result agrees nicely with that in previous literature. We explicitly verified that, while with the existing methods part of the entropy current remains undetermined, our procedure leads to a unique expression solely by using second order transport data.

We pointed out some open issues regarding the formulation of dynamical KMS transformations in the quantum regime.
There is a potential ambiguity and at the moment there is no obvious principle to fix it.

There are also other conceptual
issues in the quantum regime. For example, let us consider a neutral conformal fluid whose only scale is then the local
inverse temperature $\beta$ which provides the UV cutoff for the hydrodynamic effective action. This is, however, also
the typical scale of quantum fluctuations. While one can treat quantum effects perturbatively to maintain locality,
it appears that there is no separation of scales and thus not clear whether effective field theory approach still makes sense at all
in the full quantum regime.

\vspace{0.2in}   \centerline{\bf{Acknowledgements}} \vspace{0.2in}
We thank  P. Gao, S.~Rajagopal and D. T. Son  for discussions. Work supported in part by funds provided by the U.S. Department of Energy
(D.O.E.) under cooperative research agreement DE-FG0205ER41360.

\appendix

\section{A simple argument} \label{app:pro}

For this purpose,  we first note a general result regarding an on-shell action:
suppose an action has a symmetry
\be \label{1sym}
I [\chi; \phi] = I [\tilde \chi; \tilde \phi]
\ee
where variables with a tilde are related to original variables by some transformation,
then
\be \label{1onsh}
I_{\rm on-shell}  [\phi] = I_{\rm on-shell}  [\tilde \phi] \ .
\ee
To see this, note equation~\eqref{1sym} implies that
\be
\tilde \chi^{\rm cl} [\phi] = \chi^{\rm cl} [\tilde \phi] \ ,
\ee
and thus
\be
I_{\rm on-shell}  [\phi]  = I [\chi^{\rm cl} [\phi]; \phi] = I [\tilde \chi^{\rm cl} [\phi]; \tilde \phi] =
I [\chi^{\rm cl} [\tilde  \phi]; \tilde \phi]  = I_{\rm on-shell}  [\tilde \phi] \ .
\ee

\section{Absorbing $V^\mu$ by total derivatives}\label{app:totd}

Dynamical KMS invariance requires
\be
\label{LLVV}
\tilde{\mathcal L}-\mathcal L=\p_\mu V^\mu \ .
\ee
From $Z_2$ nature of the dynamical KMS transformation, acting on~\eqref{LLVV} with another dynamical KMS transform we find
\be
 \mathcal L-\tilde{\mathcal L}=-\p_\mu \tilde V^\mu \ ,
 \ee
where we used that $\widetilde{\p_\mu V^\mu}=-\p_\mu \hat V^\mu$, where $\tilde V^\mu$ is the dynamical KMS transform of $V^\mu$. We then find
\be\p_\mu V^\mu = \p_\mu \tilde  V^\mu
\ .\ee
Splitting $V^\mu_0$ from $V^\mu$, i.e.  $V^\mu=i V_0^\mu+V_a^\mu$, we can write (\ref{LLVV}) as
\be
\tilde{\mathcal L}-\mathcal L=i \p_\mu V_0^\mu+\frac 12(\p_\mu V_a^\mu+\p_\mu \tilde  V_a^\mu)=i \p_\mu V_0^\mu+\frac 12(\p_\mu V_a^\mu-\widetilde{\p_\mu  V_a^\mu}) \ .
\ee
Now redefining $\sL \to \sL - \frac 12 \p_\mu V_a^\mu$
we then find that
\be
 \tilde{\mathcal L} -\mathcal L = i \p_\mu V_0^\mu
\ee
i.e. all $V^\mu_k$ in~\eqref{vv} with $k \geq 1$ can be set to zero by shifting $\sL$ by a total derivative.

\section{A special basis} \label{app:dec}

Consider a generic tensor $T^{\mu_1\cdots \mu_k}$ which is an $n$th derivative order expression of $u^\mu,\ T$ and $\mu$. This can be expanded in terms of a list of independent order $n$ tensors $u^{\mu_1\cdots\mu_k}_{(n)1},\dots,\ u^{\mu_1\cdots\mu_k}_{(n)m}$, i.e.
\be\label{expan} T^{\mu_1\cdots \mu_k}=\sum_{i=1}^m c_i(\tau,\mu)\,u^{\mu_1\cdots\mu_k}_{(n)i},\ee
where $c_i(\tau,\mu)$ are functions of $\tau$ and $\mu$, and where (\ref{expan}) can be seen as a generalization of (\ref{1p})-(\ref{11p}). In the reminder we shall show that the list of the $u^{\mu_1\cdots\mu_k}_{(n)i}$'s can be rearranged into a list constituted by $v^{\mu_1\cdots\mu_k}_{(n)1},\dots,\ v^{\mu_1\cdots\mu_k}_{(n)p}$ and $w^{\mu_1\cdots\mu_k}_{(n)1},\dots,\ w^{\mu_1\cdots\mu_k}_{(n)q}$, with $p+q=m$, such that the $w^{\mu_1\cdots\mu_k}_{(n)i}$'s contain at least one factor of $E_\mu,\ E_1$ and $E_2$ (defined in (\ref{0eom})), or one factor of derivatives acting on them, and such that the $v^{\mu_1\cdots\mu_k}_{(n)i}$ are not related through the ideal equations of motion $E_\al=0$, i.e. it is not possible to find functions $c_i(\tau,\mu)$ and a tensor $w^{\mu_1\cdots\mu_k}$ such that, for some $v^{\mu_1\cdots\mu_k}_{(n)j}$,
\be v^{\mu_1\cdots\mu_k}_j=\sum_{i\neq j}^p c_i v^{\mu_1\cdots\mu_k}_{(n)i}+w^{\mu_1\cdots\mu_k} \ . \ee
For ease of notation, in the reminder we shall drop the space-time indices $\mu_1\cdots \mu_k$ and the subscript $(n)$.

First we choose a list of tensors $v_1,\dots,\ v_p$ among $u_1,\dots,\ u_m$ that are independent after imposing $E_\al=0$, with $p\leq m$. Up to permutations, we can assume that $v_1=u_1$, $\dots$, $v_p=u_p$. This clearly implies that $u_{p+1},\dots, u_m$ are generated by $v_1,\dots,\ v_p$ upon using $E_\al=0$, i.e. there are functions $c_{ij}(\tau,\mu)$ such that
\be\label{uw1} u_i=\sum_{j=1}^p c_{ij}v_j+w_{i-p},\quad i=p+1,\dots,m,\ee
where $w_i=0$ after setting $E_\al=0$, for $i=1,\dots,q$, and $q=m-p$. The latter property implies that $w_i$ is proportional to either of $E_\mu,\ E_1$ or $E_2$, or to derivatives acting on them. Note that the $w_i$'s are independent from each other and from the $v_i$'s. To see this, assume by contradiction that, for some $w_i$, with $1\leq i\leq q$, there are $d_j(\tau,\mu)$ and $e_j(\tau,\mu)$ such that
\be w_i= \sum_{j=1}^p d_j v_j+\sum_{\substack{j=1 \\j\neq i}}^q e_j w_j,\ee
then, from (\ref{uw1}),
\be\begin{split} u_{i+p}=&\sum_{j=1}^p c_{i+p,j}v_j+\sum_{j=1}^p d_j v_j+\sum_{\substack{j=1 \\j\neq i}}^q e_j \left(u_{j+p}-\sum_{k=1}^p c_{j+p,k}v_k\right)\\
=&\sum_{j=1}^p\left(c_{i+p,j}+d_j-\sum_{\substack{k=1 \\k\neq i}}^q e_k c_{k+p,j}\right)u_j+\sum_{\substack{j=1 \\j\neq i}}^q e_j u_{j+p},\end{split}\ee
which cannot happen, as $u_i$ is independent from $u_k$, with $k\neq i$. This concludes the proof.

\section{Action for conformal fluids in fluid spacetime } \label{app:conf1}

In this Appendix, we elaborate more on the explicit form of the conformal charged fluid action
in fluid spacetime without taking $\hbar \to 0$ limit.

The discussion parallels that of a general fluid in Sec. \ref{sec:2a}  so we will only highlight the differences.

Using definition (\ref{hnew}), we decompose $\hat h_{sab} = e^{- 2 \tau_s} h_{s ab}$ as
\be \label{0nep0}
\hat h_{sab} d\sig^a d\sig^b =  - E^2_s  \le( d\sig^0 -  v_{si} d\sig^i \ri)^2 + %e^{2  \eta}
\al_{sij} d\sig^i d\sig^j, \\ %= - \ell_a \ell_b + \De_{ab}
\ee
with
\be \label{mhat}
E_s = e^{- \tau_s} b_s ,  \qquad \al_{sij} = %e^{-2 \tau} \lam_i{^\mu} \lam_{j \mu}\equiv
e^{-2 \tau_s} a_{sij}  \ . %, \quad \hmu = e^\tau \mu = e^{\tau} u^\mu A_{\mu} + D_0 \vp,  %u^\mu B_\mu ,
%\quad \fb_i =   \lam_{i}{^\mu}  A_{\mu} + D_{i} \vp,   %\lam_i{^\mu} B_\mu , \qquad
%\hat A_\mu = A_\mu - \p_\mu \vp
\ee
Instead of~\eqref{decB}  it is more convenient to decompose $B_{sa}$ as
\be
B_{sa} d\sig^a = \hmu_s E_s (d\sig^0 - v_{si} d\sig^i) + \fb_{si} d\sig^i
\ee
with $\hmu_s = e^{\tau_s} \mu_s$ and the corresponding symmetric and antisymmetric combinations
$\hmu_r = \ha (\hmu_1 + \hmu_2)$ and $\hmu_a = \hmu_1 - \hmu_2$.
Instead of~\eqref{1nr}--\eqref{1na}, $E_{r,a}, \chi_a$ are now defined as
\begin{gather}  \label{11nr}
 E_r =\ha (E_1 + E_2) =
 \ha \le(e^{- \tau_1} b_1 + e^{-\tau_2} b_2 \ri) ,\quad E_a = \log (E_2^{-1} E_1) =  - \tau_a + \log \le(b_2^{-1} b_1 \ri) , \\
 \al_{r ij} = \ha (\al_{1ij} + \al_{2ij}),
  \qquad \chi_a = \ha \log \det (\al_2^{-1} \al_1) \ .
  \label{11na}
\end{gather}
We also have $\tau_{r,a}$  introduced in Sec.~\ref{sec:7a}, and $\Xi, v_{ai}, v_{ri}, \fb_{ai}, \fb_{ri}$
are the same as those in Sec.~\ref{sec:2a}.

We will now use $\al_r$ to raise and lower $i,j$ indices.
The time covariant derivative is the same as~\eqref{tder} except that one should use $E_r$ defined as in~\eqref{11nr}.
 and the spatial covariant derivatives
are the same as~\eqref{der}--\eqref{dest} except that $\tilde \Gamma^i_{jk}$  should be replaced by that associated with
$\al_{rij}$, more explicitly,
 \be \label{1omm}
\tilde\Gamma^i_{jk} \equiv \ha \alpha^{il}_r \left(d_j \alpha_{rkl}+ d_k \alpha_{rjl}
-d_l\alpha_{rjk}\right)=\Gamma^i_{jk}+\frac{1}{2}\alpha^{il}_r \left(v_{rj}\partial_0\alpha_{rkl} + v_{rk}\partial_0\alpha_{rjl}
-v_{rl}\partial_0\alpha_{rjk}\right)
\ee
with $\Gamma^i_{jk}$ the Christoffel symbol corresponding to $\al_r$.

We can write the action as
\be  \label{1intm}
\int d^d \sig \, \sqrt{\al_r} E_r \, \sL
\ee
where the integration measure $\sqrt{\al_r} E_r$ differs from that of~\eqref{intm} by some factors of $\tau$.
Note that equation~\eqref{cftf} implies that there cannot be any dependence on $\tau_r, \tau_a$ other than those included
in $\hat h_{sab}$.  $\sL^{(1)}$ to first derivative order and $\sL^{(2)}$ to zeroth derivative order are given by~\eqref{kee}--\eqref{f3} and~\eqref{aa0} that one should eliminate all terms which involve $\tau$ explicitly and all coefficients
are functions of $\hmu_r$ only.

We now write down $\sL^{(1)}$ to second derivative order for neutral conformal fluids. For this purpose we need to use the
the torsion $\ft_{ij}$ and various curvatures
introduced  in Sec. V A3 of~\cite{CGL} except that they should be accordingly modified as those associated with $\al_r$.
The torsion $\ft_{i j}$ is defined by
\be\label{1tor}
[D_i , D_j ] \phi \equiv \ft_{ij} D_0\phi, \qquad \ft_{ij} = E_r (d_i v_{rj} - d_j v_{ri}) ,
  \ee
and  the ``Riemann tensor'' $\tilde R^k_{\ lij}$ by
\be\label{1defR}
[D_i,D_j]\phi_k=\tilde R_{ijk}{^l}\phi_l+ \ft_{ij} D_0\phi_k \
\ee
with
\be \label{1riem}
\tilde R_{ijk}{^l} = d_j \tilde\Gamma^l_{ik} - d_i \tilde\Gamma^l_{jk}  + \tilde\Gamma^m_{ki}\tilde\Gamma^l_{jm} - \tilde\Gamma^m_{kj}\tilde\Gamma^l_{im} \ .
\ee
Note that
\bega \label{2defR}
\tilde R_{ijkl} + \tilde R_{ijlk} = - \ft_{ij} D_0 \al_{rkl}.
\end{gather}
and as a result, there are two ``Ricci tensors'':
\be
\tilde R_{ik}^1 =  \tilde R_{ijk}{^j}, \qquad \tilde R_{ik}^2 =  \tilde R_{ij}{^j}{_k},
\ee
neither of which is symmetric. We also introduce
\be
W_{ik} = \tilde R_{ik}^1 + \tilde R_{ik}^2 = - \ft_{ij} \al_r^{jl} D_0 \al_{r kl} ,
\qquad S_{ik} = \ha \le(\tilde R_{ik}^1 - \tilde R_{ik}^2 \ri) \ ,
\ee
where the second equality of the first equation follows from~\eqref{2defR}.

Now at second derivative order $\sL^{(1)}$ can be written as
\bega \label{1kee}
\sL^{(1)}_2 = f_1 E_a + f_2  \chi_a + \lam_{21} V_{a}^i \al^{jk} D_j D_0 \al_{ik} + \lam_{22} V_a^i\alpha^{jk} D_j \mathfrak t_{ik}
\cr
-  \Xi^{ij} \le[  {\tilde \eta  \ov 2}  D_0 \al_{rij}  + \eta_1 D_0^2 \al_{ij} + \eta_2 D_0 \al_{ik} D_0 \al_{jl} \al^{kl}+ \eta_3 S_{ij}+\eta_4 \alpha^{kl}\mathfrak t_{ki} \mathfrak t_{jl}+\eta_5 W_{ij}
 \ri],
 %\cr
%+  V_{a}^i  \le[\lam_{3} D_i E_r  + \lam_{33} D_0 D_i E_r  + \lam_{34} \al^{jk} D_j E_r D_0 \al_{ik}
%+ \lam_{35} D_i \tr (\al^{-1} D_0 \al) + \lam_{36} \al^{jk} D_j D_0 \al_{ik} \ri] + \cdots
\end{gather}
where
\be \label{2kee}
f_1 =  f_{11} %+  \ha f_{14}  \tr (\al^{-1} D_0 \al) + f_{15} {\rm tr}^2 (\al^{-1} D_0 \al)
+ f_{15} \tr\le(D_0 \al \al^{-1} D_0 \al \al^{-1}\ri)  %\cr&& \quad
 + f_{16} \tr\le(\al^{-1} D_0^2 \al \ri)
%+ f_{18} \al^{ij} D_i E_r D_j E_r + f_{19} D^i D_i E_r
+ f_{17} S_{ij} \al_r^{ij} + f_{18} W_{ij} \al_r^{ij} +  f_{19}\frak{t}^{ij}\frak{t}_{ij} ,  %\cdots
% , \\
%f_2 &=&  f_{21} + f_{22} D_0 \tau_r + f_{23} D_0 \hmu_r +  {f_{24} \ov 2} \tr (\al_r^{-1} D_0 \al_r)
%+ {\rm higer \, derivatives} ,
%\tilde \eta &= & \eta_1 + \eta_2 \tr (\al^{-1} D_0 \al) + \cdots \\
%\lam_3 &= & \lam_{31} + \lam_{32} \tr (\al^{-1} D_0 \al)  + \cdots
\ee
and $f_2$ has the same structure as $f_1$. In~\eqref{1kee}--\eqref{2kee}, all coefficients are constants, and we have dropped terms which vanish on the zeroth order equations of motion:
\be
\Tr \le(\alpha_r^{-1} D_0\alpha_{r}\ri)= 0, \qquad  D_i E_r =0 \ .
\ee

The $\hbar\to 0$ limit of the second line of (\ref{1kee}) gives the second order stress tensor (\ref{Tmn1}) with the correspondence
\be \begin{gathered}
\eta_1=-\frac 18(2 f_5+f_1),\quad \eta_2=\frac 1{32}(8 f_5-4 f_2+7f_1),\\
\eta_3=-\frac 12 f_1,\quad \eta_4=\frac 1{32}(4f_3+11f_1),\quad \eta_5=-\frac 18 (2f_5+f_4+f_1)\ .
\end{gathered}\ee

\section{Useful identities}\label{app:id}

In this Appendix we present the identities  used in Sec.~\ref{sec:conf}. From
\be [ \hat\nabla_\mu,\hat\nabla_\nu]\beta_\alpha = \hat R_{\mu\nu\alpha}^{\ \ \ \ \beta}\beta_\beta,\ee
and contracting with $ \hat \Delta_{\mu\nu}$ or $\beta_\mu$, we find
\begin{align}
\hat R_{\mu\nu}\beta^\mu\beta^\nu &=-\frac 14(\hat \sigma^2-\hat\omega^2)\label{f22}\\
\hat\p\hat\sigma_{\mu\nu}&= 2 \hat R_{\alpha\mu\nu\beta} \beta^\alpha \beta^\beta-\frac 12(\hat \sigma_\mu^\alpha\hat\sigma_{\alpha\nu}+\hat\omega_\mu^{\ \alpha}\hat\omega_{\alpha\nu})\\
\label{f44}\hat\p\hat\omega_{\mu\nu}&=\hat\sigma_{[\mu}^{\ \ \alpha}\hat\omega_{\nu]\alpha}\\
\hat\Delta_\alpha^\mu\hat\nabla_\mu\hat\nabla_\beta\hat\sigma^{\alpha\beta}&= 2\hat\Delta^{\mu\alpha}\hat\nabla_\mu(\hat R_{\alpha\beta}\beta^\beta)-\frac 12\hat\p(\hat\omega^2),
\end{align}
where $\hat R_{\mu\nu}=\hat R_{\mu\alpha\nu}^{\ \ \ \ \alpha}$, and where we used that $\hat\nabla_\mu\beta^\mu=\beta^\mu\hat\nabla_\mu\beta^\alpha=0$. We also have the Bianchi identity $\hat\nabla_\mu\left(\hat R^{\mu\nu}-\frac 12 \hat g^{\mu\nu}\hat R\right)=0$.

\end{document}